\documentclass[a4paper,11pt]{article}
\usepackage[utf8]{inputenc}
\usepackage{amsfonts}
\usepackage{amssymb}
\usepackage{graphics}
\usepackage{fancyhdr}
\usepackage{fancyvrb}
\usepackage{fancybox}
\usepackage{float}
\usepackage{geometry}
\usepackage{minitoc}
 \usepackage{indentfirst}
 \usepackage{multicol}
 \usepackage[outerbars]{changebar}
 \usepackage{pifont,textcomp}
 \usepackage{amsfonts}
 \usepackage{amsmath}
 \usepackage{amscd}
 \usepackage{array}
\usepackage{multirow}
 \usepackage{mathrsfs}
 \usepackage{amssymb}
 \usepackage{amsthm}

\title{ Quantization and coherent states for a time-dependent Landau problem}
\title{ Quantization and coherent states for a time-dependent Landau problem}
\author{ Lat\'evi Mohamed Lawson\\ 
	\space\\
	 ${}^{a}\!$ Universit\'{e} de Lom\'{e}, Facult\'{e} des Sciences, D\'{e}partement de Physique,\\
	 Laboratoire de Physique des Mat\'{e}riaux et des Composants\\
	 à Semi-Conducteurs, 01 BP 1515 Lom\'{e}, Togo\\\\
	 ${}^{b}\!$ Institut de Mathématiques et de Sciences Physiques (IMSP),\\ Laboratoire de Recherche en Physique Théorique (LRPT).\\
	 01 BP 613 Porto-Novo, Rep. du Bénin\\\\
 latevi@aims.edu.gh }

\begin{document}
\maketitle

	
   The ordinary Landau problem   consists of describing a charged particle in time-independent magnetic field.
   In the present case the problem is generalized onto time-dependent uniform electric fields with time-dependent mass and harmonic frequency \cite{102}.
   The spectrum of a Hamiltonian describing this system is obtained. The configuration space wave functions of the model is expressed in terms of the
   generalised Laguerre polynomials. To diagonalize the time-dependent Hamiltonian we employ the
   Lewis-Riesenfeld method of invariants. To this end, we introduce an unitary transformation in the
   framework of the algebraic formalism to construct the invariant operator of the system and then
   to obtain the exact solution of the Hamiltonian. We recover the solutions of the ordinary Landau
   problem in the absence of the electric and harmonic fields, for a constant particle mass. The quantization of this system 
   exhibits many  symmetries
   such  as $U(1), SU(2), SU(1,1)$. We therefore construct the coresponding coherent states  and the associated photon added and nonlinear coherent states.



\section{The model}

We consider the problem of a charged particle of charge q with time-dependent mass $M(t)$
and frequency $\omega(t)$ moving in a two-dimensional plane and subjected to a static magnetic $B$ field
perpendicular to the plane and a spatially uniform but time-dependent electric field $E(t)$
in lying position in the Euclidean plane. In the symmetric gauge, the vector potential and the time-dependent
scalar potential are given by $A_i(x i )=-\frac{1}{2}B\epsilon_{ij}x_j$ and $\varphi(t)=E_i(t)x_i $, respectively (with $i,j= 1,2$).
The Lagrangian of the system is written as follows
\begin{eqnarray}
	L(x_1,x_2,\dot{x}_1,\dot{x}_2,t)&=&\frac{1}{2}M(t)(\dot x_1^2+\dot x_2^2)-\frac{1}{2}M(t)\omega^2 (t)(x_1^2+x_2^2)\cr&&
	+q\left[A_{x_1}(x_1,x_2) \dot x_1+A_{x_2}(x_1,x_2) \dot x_2\right]-q\varphi(x_1,x_2,t).
\end{eqnarray}
The Euler-Lagrange equations of motion for the
system are
\begin{equation}\label{z}
	\begin{cases}
		\ddot{x}_1 -\frac{qB}{M(t)}\dot{x}_2+\omega(t) x_1-\frac{q}{M(t)}E_{x_1}(t)=0, \\
		\ddot{x}_2 +\frac{qB}{M(t)}\dot{x}_1+\omega(t) x_2- \frac{q}{M(t)} E_{x_2}(t)=0.
	\end{cases}
\end{equation}
Since we are in two dimensional configuration space, we can look for the solutions of the classical equations  in the complex system by
setting $z=x_2+ix_1$. The classical  equation of motion  in term of the coordinate $z$  is 
\begin{eqnarray}
	\ddot{z}+i\omega_c(t)\dot{z}+\omega(t)z=E_0(t), 
\end{eqnarray}
where $\omega_c(t)=\frac{qB}{M(t)}$ and $E_0=\frac{qE_{x_1}}{M(t)}+i\frac{qE_{x_2}}{M(t)}$. The general homogenous  solution
of this equation is 
\begin{eqnarray}
	z(t)=A e^{-i\omega_+t}+B e^{i\omega_-t}+\frac{E_0}{\omega^2},
\end{eqnarray}
where $\omega_+=\sqrt{\omega^2+\frac{1}{4}\omega_c^2}+\frac{\omega_c}{2}$ and 
$\omega_-=\sqrt{\omega^2+\frac{1}{4}\omega_c^2}-\frac{\omega_c}{2}$ .

The canonical momentum associated with the variables $x_1$ and $x_2$ are
\begin{equation}\label{w}
	\begin{cases}
		p_1=\frac{\partial L}{\partial \dot{x}_1}= M(t) \dot{x}_1+qA_{x_1} \\
		p_2=\frac{\partial L}{\partial \dot{x}_2}= M(t) \dot{x}_2+qA_{x_2}.
	\end{cases}
\end{equation}
The  Hamiltonian is given by
\begin{eqnarray} \label{lan23}
	H(x_1,x_2,p_1,p_2,t)&=& \dot{x}_1 p_1+\dot{x}_2 p_2-L\cr
	&=& \frac{1}{2M(t)}\left[p_1+ \frac{q}{2}  Bx_2\right]^2+\frac{1}{2M(t)}\left[p_2-\frac{q}{2}  Bx_1\right]^2\cr
	&& +\frac{1}{2} M(t) \omega^2(t)(x_1^2+x_2^2) -q[E_1(t)x_1+E_2(t)x_2].
\end{eqnarray}
Note that for $\vec E(t)=\vec 0$, $\omega(t)=0$  and for a constant mass $M$, we obtain a Hamiltonian operator
describing the  ordinary Landau problem \cite{178,179,180}.

We introduce the following changes of variables 
\begin{eqnarray}
	x&=&x_1+\frac{qE_1(t)}{M(t)\omega^2(t)},\,\,\, y=x_2+\frac{qE_2(t)}{M(t)\omega^2(t)},\label{lan18}\\
	p_x&=&p_1-\frac{q^2BE_2(t)}{2M(t)\omega^2(t)},\,\,\, p_y=p_2-\frac{q^2BE_1(t)}{2M(t)\omega^2(t)},\label{lan19}
\end{eqnarray}
and get the transformed  Hamiltonian  in its simple  form rewritten as follows
\begin{equation}
	H(t)=\frac{1}{2M(t)}(p_x^2+p_y^2)+\frac{\Omega^2(t)M(t)}{2}(x^2+y^2)-\frac{\omega_c (t)}{2}L_z-\frac{q^2E^2(t)}{2M(t)\omega(t)},
\end{equation}
where $L_z=xp_y-yp_x$ is the angular  momentum of  the system,  $\omega_c(t)=\frac{qB}{M(t)}$ is the cyclotronic frequency of oscillations,
$\Omega(t)=\sqrt{\omega^2(t)+\frac{1}{4}\omega_c^2(t)}$ is the general frequency of oscillations and
$ E(t)=\sqrt{E_1^2(t)+E_2^2(t)}$. 
From the above  choice of the gauge symmetric, the  system generated the angular momentum $L_z$ that commutes 
with the Hamiltonian of the system
\begin{eqnarray}
	[H(t),L_z]=0.
\end{eqnarray}
Therefore, both operators  admit a  commun basis in which they can be simultaneously diagonalized.
Furthermore, $L_z$  can be interpreted as the generator of  $SO(2)=U(1)$ group which  conserve the invariance of the orientation and the rotation in the plane
i.e the eigenstates of the Hamiltonian  $H(t)$ is invariant under this rotation. The group of rotation
of the system around the $z$-axis is written  as  follows
\begin{eqnarray}
	R_z(\alpha)=\exp\left(-\frac{i\alpha}{\hbar} L_z \right)=\sum_{n=0}^\infty \frac{1}{n!} \left(-\frac{i\alpha}{\hbar}L_z\right)^n,
\end{eqnarray}
with $\alpha\in \mathbb{R}$ is the angle of rotation. Since  the magnetic field is constant and uniform along the $z$-axis this group satisfy
the commutation relation
\begin{eqnarray}
	R_z(\alpha) R_z(\beta)= R_z(\beta) R_z(\alpha),\,\,\,\,\,\,\alpha,\beta\in\mathbb{R}.
\end{eqnarray}
Furthermore, the operator $L_z$ is self-adjoint then the condition of unitarity is guarantee. In fact the quantum 
rotational operator  obeys the following important properties
\begin{eqnarray}
	R_z(\alpha_3,\alpha_2)R_z(\alpha_2,\alpha_1)&=&R_z(\alpha_3,\alpha_1),\\
	R_z^\dag(\alpha_2,\alpha_1)=R_z^{-1}(\alpha_2,\alpha_1)&=&R_z(\alpha_1,\alpha_2),\\
	R_z(0)&=&\mathbb{I}.
\end{eqnarray}
Since the system of our study  contained parameters  which change in time,  we  consider suitable Hilbert space 
$\mathcal{H}= L^2(\mathbb{R}^3)$ of square integrable wavefunctions $\psi(x,y)$ over $\mathbb{R}^2$ and
the time-parameter $t\in \mathbb{R}$. For simplicity we  consider the wavefunction in the form $\psi(x,y,t)$ defined over $ L^2(\mathbb{R}^3)$.\
The Schr\"{o}dinger  equation  is  explicitly  written as  follows
\begin{eqnarray}\label{lan2}
	i\partial_t\psi (x,y,t)&=&-\frac{1}{2M(t)}(\partial_x^2+\partial_y^2)\psi(x,y,t)+\frac{\Omega^2(t)M(t)}{2}(x^2+y^2)\psi(x,y,t)
	\cr 
	&&+i\frac{\omega_c(t)}{2}(x\partial_y-y\partial_x)\psi(x,y,t)-\frac{q^2E^2(t)}{2M(t)\omega(t)}\psi(x,y,t),  
\end{eqnarray}
where  we set $\hbar=1$.

Since the Hamiltonian  involves time-dependent parameters, the above differential equation cannot be easily  solved.
Then, we use the so-called Lewis and Riesenfeld method  \cite{47} which consists in constructing  a Hermitian operator $I(t)$ 
that fulfills the condition
\begin{equation}\label{lan3}
	\frac{dI}{dt}=\frac{\partial I}{\partial t}+\frac{1}{i}[I,H]=0.
\end{equation}
The Schr\"{o}dinger equation for the invariant operator reads as
\begin{equation}\label{lan1}
	I(t)\phi(x,y,t)=E\phi(x,y,t),\,\,\,\text{with}\,\,\,\,\, \frac{\partial E}{\partial t}=0,\,\,\,\phi(x,y,t)\in \mathcal{H}.
\end{equation}
The solutions of Schr\"{o}dinger  equation  (\ref{lan1}) is linked to the invariant's eigenfunction by
\begin{equation}
	\psi(x,y,t)=e^{i\gamma(t)}\phi(x,y,t),
\end{equation}
where  the time-dependent Lewis-Riesenfeld phase $\gamma(t)$ satisfies the following equation
\begin{equation}\label{lan14}
	\frac{d\gamma (t)}{dt}=\langle \phi|i\frac{\partial}{\partial t}-H(t)|\phi\rangle.
\end{equation}
Therefore to determine the solution of the system (\ref{lan2}), it found important to construct firstly the Hermitian invariant $I(t)$
and secondly determine its eigensystems.
\subsection{The invariant operator}
We look for the Hermitian operator in the form
\begin{equation}
	I(t)=\frac{1}{2}\left[\alpha(t) (x^2+y^2)+\beta(t)(p_x^2+p_y^2)+\delta(t)(xp_x+p_xx+yp_y+p_yy)\right],
\end{equation}
where $\alpha, \beta,$ and $\delta$ are real coefficient functions. Using (\ref{lan3}) and the following equations
\begin{eqnarray}
	[x^2,p_x^2]&=&2i\{x,p_x\},\,\,\,\,[x^2,\{x,p_x\}]=4ix^2,\,\,\,[p_x^2,\{x,p_x\}]=-4ip_x^2,
\end{eqnarray}
\begin{eqnarray}
	[y^2,p_y^2]&=&2i\{y,p_y\},\,\,\,[y^2,\{y,p_y\}]=4iy^2,\,\,\,[p_y^2,\{y,p_y\}]=-4ip_y^2.
\end{eqnarray}
We obtain the first-order linear differential equations for the unknown coefficient functions 
\begin{eqnarray}
	\dot{\alpha}- 2 M\Omega^2 \delta &=&0,\label{lan5}\\
	\dot{\beta}+\frac{2}{M}\delta&=&0,\label{lan6}\\
	\dot{\delta}+\frac{1}{M}\alpha -M\Omega \beta&=&0,\label{lan7}\\
	\dot{\delta}&=&0.\label{lan4}
\end{eqnarray}
Equation (\ref{lan4}) implies that $\delta$ is a constant. In the rest of the text, we choose
$\delta = 0$ for the other invariant. From (\ref{lan5})-(\ref{lan7}), it follows:
\begin{equation}
	\frac{d}{dt}(\delta^2-\alpha \beta)=0,
\end{equation}
whence
\begin{equation}
	\delta^2-\alpha \beta=-\kappa^2,
\end{equation}
where $\kappa$ is either a real or an imaginary constant (so that $\kappa^2$ is real).
It  is convenient to introduce another real function $\rho$ defined by
\begin{equation}
	\beta(t)=\rho^2(t).
\end{equation}
For an arbitrary positive constant $\kappa$, the other coefficients are
\begin{eqnarray}
	\delta(t)=-M\dot{\rho}\rho,\,\,\,\,\,\,\alpha(t)=\frac{\kappa^2}{\rho^2}+M^2\dot{\rho}^2.
\end{eqnarray}
The Hermitian invariant therefore acquires the form
\begin{equation}
	I(t)=\frac{1}{2}\left[\frac{\kappa^2}{\rho^2}x^2+\frac{\kappa^2}{\rho^2}y^2+\left(\rho p_x-M\dot{\rho}x\right)^2+
	\left(\rho p_y-M\dot{\rho}y\right)^2\right],
\end{equation}\label{lan8}
where the function $\rho$ is the solution of the so-called nonlinear Ermakov-Pinney equation \cite{133,134,135}
\begin{equation}\label{lan8}
	\ddot{\rho}+ \frac{\dot{M}}{M}\dot{\rho}+\Omega^2(t)\rho=\frac{\kappa^2}{M^2\rho^3}.
\end{equation}

As matter of fact, for concrete computations of measurable quantities one needs to address the auxiliary
problem and solve the equations explicitly for the time-dependent functions appearing in
the Hamiltonian. Surprisingly little attention has been paid to this problem in the context
of solving time-dependent Hamiltonian systems and therefore we will discuss the solutions
of our auxiliary equation (\ref{lan8}) in the next subsection.

\subsubsection{Solutions of the   Ermakov-Pinney equation}

The simplest special solution arises when  taking $ M(t)=\tau$=constant ($\dot{M}=0$),
consequently the friction coefficient vanishes and the solution found by Pinney \cite{133} was
\begin{equation}
	\rho(t)=\sqrt{v_1^2+ \nu^2v_2^2 \mathcal{W}^{-2}},
\end{equation}
where $\nu^2=\frac{\kappa^2}{\tau^2}$; $v_1,v_2$ are linear independent solutions of the equation
\begin{equation}
	\ddot{v}+\omega^2v=0,
\end{equation}
and $\mathcal{W}=v_1\dot{v_2}-\dot{v_1}v_2=$const $\neq 0$  is their Wronskian.

For still the non-dissipative solutions ($\dot{M}=0$), we can simply pre-select any explicit form
for $\Omega(t)=\tau e^{\alpha t} (\alpha,\tau \in \mathbb{R}$, we solve (\ref{lan8}) in terms of the Bessel functions
and subsequently obtain the particular solution by means
\begin{equation}
	\rho(t)=\sqrt{\frac{\pi^2\nu^2}{\alpha^2 A_1^2}Y_0^2\left(\frac{2\nu e^{\frac{\alpha t}{2}}}{\sqrt{\tau}\alpha}\right)+
		A_1^2 J_0^2\left(\frac{2\nu e^{\frac{\alpha t}{2}}}{\sqrt{\tau}\alpha}\right)}
\end{equation}
with integration constant $A_1\in \mathbb{R}$  and $J_0$ , $Y_0$ denoting the Bessel functions of first and
second kind, respectively.

Now, for special values of $M(t) = e^{-\alpha t}$ and $\Omega(t)=\frac{\sqrt{5}}{2}\alpha$, the transformation $\rho=e^{-\frac{1}{2}\alpha t}y$
leads to the Yermakov’s equation \cite{134}
\begin{eqnarray}
	\ddot{y}+\alpha^2 y=\kappa^2y^{-3}.
\end{eqnarray}
Using the solution of this equation, we obtain for $\rho$ the solutions:
\begin{eqnarray}
	\rho(t)= e^{-\frac{1}{2}\alpha t}\left[\frac{\kappa s^2}{d_1}+\frac{s^2}{d_2} \left(d_2+d_1\int_0^t \frac{dt'}{s^2}\right)^2\right]^{\frac{1}{2}},
\end{eqnarray}
where  $d_1, d_2$ arbitrary real constants and $s(t')$ is  a function such a
\begin{eqnarray}
	s(t')&=& e_1\sin\left(\alpha t'\right)+e_2\cos\left(\alpha t'\right),
\end{eqnarray}
where $e_1$ and $e_2$ are arbitrary real constants.

\subsubsection{Eigensystems of the invariant operator}

In order to   solve easily  the eigenvalue equation of the invariant operator, we consider  the unitary operator $U$ which is written as follows 
\begin{equation}\label{lan20}
	U=\exp\left[-i\frac{M\dot{\rho}}{2\rho}\left(x^2+y^2\right)\right],\,\,\,\,\,\, U^\dag U= UU^\dag =\mathbb{I}.
\end{equation}
This operator transforms,  the invariant operator's eigenfunction into 
\begin{equation}
	\phi'(x,y,t)=U\phi(x,y,t),
\end{equation}
reduces  the invariant operator  into the form
\begin{eqnarray}
	I'(t)=UI(t)U^\dag=\frac{1}{2}\left[\rho^2(p_x^2+p_y^2)+\frac{\kappa^2}{\rho^2}(x^2+y^2)\right],
\end{eqnarray}
but keeps its eigenvalue invariant 
\begin{eqnarray}
	I'(t)\phi'(x,y,t)&=&E_\alpha' \phi'(x,y,t)\cr
	UI(t)U^\dag U\phi(x,y,t)&=&  E_\alpha'  U\phi(x,y,t).\label{lan9}
\end{eqnarray}
Multiplying the equation (\ref{lan9}) with the operator $U^\dag$  and using the unitarian relation $U^\dag U =\mathbb{I}$,
we get
\begin{eqnarray}
	E_\alpha=E_\alpha'.
\end{eqnarray}
Instead of using   analytical method  \cite{181,182,183} to solve the eigenvalue equation  of the invariant operator, 
we  achieve its diagonalisation  through algebraic method. We introduce the reduced  lowering and raising operators given by
\begin{eqnarray}
	a_x'&=&\frac{1}{\sqrt{2\kappa}}\left(\frac{\kappa}{\rho}x+i\rho p_x\right),\,\,\,\,
	{a'}_x^\dag=\frac{1}{\sqrt{2\kappa}}\left(\frac{\kappa}{\rho}x-i\rho p_x\right),\\
	a_y'&=&\frac{1}{\sqrt{2\kappa}}\left(\frac{\kappa}{\rho}y+i\rho p_y\right),\,\,\,\,
	{a'}_y^\dag=\frac{1}{\sqrt{2\kappa}}\left(\frac{\kappa}{\rho}y-i\rho p_y\right),
\end{eqnarray}
that fulfill the commutation relations
\begin{eqnarray}
	[a_x',{a'}_x^\dag]=\mathbb{I}=  [a_y',{a'}_y^\dag],\,\,\,\,\,[a_x',a_y']= 0= [a_x',{a'}_y^\dag].
\end{eqnarray}
Let us consider any  nonnegative integers $n_x,n_y$ and $|\phi'_{n_x,n_y}(t)\rangle$ the orthonormalized
Fock space  such as
\begin{eqnarray}
	|\phi'_{n_x,n_y}(t)\rangle&=&\frac{1}{\sqrt{n_x!n_y!}}\left({a'_x}^\dag\right)^{n_x} \left({a'_y}^\dag\right)^{n_y}|\phi'_{0,0}(t)\rangle,\label{lan10}\\
	\langle \phi'_{n_x,n_y}(t)|\phi'_{m_x,m_y}(t)\rangle&=&\delta_{n_x,m_y}\delta_{n_x,m_y},
\end{eqnarray}
with $|\psi'_{0,0}(t)\rangle$ is a normalized state annihilated  by $a'_x,a'_y$.

In order to determine the exact solution $\psi_{n_x,n_y}(x,y,t)$ of the invariant operator $I(t)$, we first express the  
ground state $|\psi_{0,0}(t)\rangle$ in the configuration space base as follows 
\begin{eqnarray}
	\phi_{0,0}(x,y,t)&=&U^\dag\langle x|\phi_0'(t)\rangle\langle y|\phi_0'(t)\rangle\cr
	&=&\left(\frac{\kappa}{\pi\rho^2}\right)^{\frac{1}{2}}\exp\left[ \left(iM\frac{\dot{\rho}}{\rho}-\frac{\kappa}{\rho^2}\right)
	\left (\frac{x^2+y^2}{2}\right)\right].
\end{eqnarray}
Then, the $nth$ eigenfunctions are obtained from (\ref{lan10}) as 
\begin{eqnarray}
	\phi_{n_x,n_y}(x,y,t)&=& U^\dag \phi_{n_x,n_y}'(x,y,t)\cr
	&=&\frac{1}{\rho}\left(\frac{\kappa}{2^{n_x+n_y}\pi\, n_x!n_y!}\right)^{\frac{1}{2}}
	H_{n_x}\left(x\frac{\sqrt{\kappa}}{\rho}\right) 
	H_{n_y}\left(y\frac{\sqrt{\kappa}}{\rho}\right)\cr&&\times
	\exp\left[\left(iM\frac{\dot{\rho}}{\rho}-\frac{\kappa}{\rho^2}\right)\left(\frac{x^2}{2}
	+\frac{y^2}{2}\right)\right],
\end{eqnarray}
where $H_{n_x}$ and $H_{n_y}$ are the Hermite polynomial of order $n_x$ and $n_y$.
To obtain the eigenvalues $E_{n_x,n_y}$ of the invariant operator $I(t)$, let us introduce a
new pair of raising and lowing  operators define as
\begin{eqnarray}
	a_j&=&U^\dag a'_j U=\frac{1}{\sqrt{2\kappa}}\left(M\dot{\rho}x_j-\rho p_j+i\frac{\kappa}{\rho}x_j\right)\label{x7},\\
	a_j^\dag&=&U^\dag {a'}_j^\dag U=\frac{1}{\sqrt{2\kappa}}\left(M\dot{\rho}x_j-\rho p_j-i\frac{\kappa}{\rho}x_j\right) \label{x8}.
\end{eqnarray}
with $j=x,y$. In term of these operators the  operator  $I(t)$ takes the form
\begin{eqnarray}
	I(t)&=& \kappa\left(a_x^\dag a_x+a_y^\dag a_y+\mathbf{I}\right),
\end{eqnarray}
The action of $a_j$ and $a_j^\dag$ on $ |\psi_{n_j}(t)\rangle$  finds expression in 
\begin{eqnarray}
	a_j^\dag|\phi_{n_j}(t)\rangle&=&\sqrt{n_j+1}|\phi_{n_j+1}(t)\rangle,\\
	a_j|\phi_{n_j}(t)\rangle&=& \sqrt{n_j}|\phi_{n_j-1}(t)\rangle,\\
	a_j^\dag a_j|\phi_{n_j}(t)\rangle&=& n_j|\phi_{n_j}(t)\rangle.
\end{eqnarray}
Basing on these  definitions, the invariant is diagonalized as follows
\begin{eqnarray}
	I(t)|\phi_{n_x,n_y}(t)\rangle&=& E_{n_x,n_y} |\phi_{n_x,n_y}(t)\rangle,\,\,\,\,\,\, E_{n_x,n_y}=\kappa\left(n_x+n_y+1\right).
\end{eqnarray}
However, as we pointed out in the previous section,  this system possesses a conserved angular-momentum
\begin{eqnarray}\label{w2}
	L_z =i(a_y^\dag a_x-a_y^\dag a_x),
\end{eqnarray}
which commutes with the invariant operator $I(t)$
\begin{eqnarray}
	[I(t),L_z]=0.
\end{eqnarray}
Although the operator $L_z$  commutes with both $ I(t)$ and $ H(t)$, but the bases $|\phi_{n_x,n_y}(t)\rangle$  does not diagonalize them simultaneously.
It convenient to work in another bases of Hilbert space  which take  account of the invariance of rotation and  diagonalizes these   operators.

\section{ U(1) symmetry and its coherent states}
\subsection{U(1) symmetry}

In order to make explicit the $SO(2)=U(1)$ circular symmetry of the system and to determine the corresponding  bases which 
can diagonalize simultaneously the invariant operator, the angular momentum
and the Hamiltonian of the system.  Let us consider  the reduced helicity  Fock algebra generators as follows
\begin{eqnarray}\label{lan11}
	a_{\pm}'&=&\frac{1}{\sqrt{2}}\left(a_x'\pm ia_y'\right),\,\,\, a_{\pm}'^\dag=\frac{1}{\sqrt{2}}\left(a_x'^\dag\mp ia_y'^\dag\right),
\end{eqnarray}
with 
\begin{eqnarray}
	[a_\pm',a_\pm'^\dag]=\mathbb{I},\,\,\,\,\,\, [a_\pm',a_\mp'^\dag]=0.
\end{eqnarray}
The inverse relations are,
\begin{eqnarray}
	a_x'&=& \frac{1}{\sqrt{2}}\left(a_+'+a_-'\right) ,\,\,\, a_x'^\dag=\frac{1}{\sqrt{2}}\left(a_+'^\dag+ a_-'^\dag\right),\\
	a_y'&=& -\frac{i}{\sqrt{2}}\left(a_+'-a_-'\right) ,\,\,\,  a_y'^\dag=\frac{i}{\sqrt{2}}\left(a_+'^\dag- a_-'^\dag\right).
\end{eqnarray}
The associated helicity-like bases
$|\phi_{n_+,n_-}'(t)\rangle$  are defined as follows
\begin{eqnarray}\label{v2}
	|\phi_{n_+,n_-}'(t)\rangle&=&\frac{1}{\sqrt{n_+!n_-!}}\left(a_+'^\dag\right)^{n_+} \left(a_-'^\dag\right)^{n_-}|\phi_{0,0}'(t)\rangle,\label{lan13}\\
	\langle \phi_{n_+,n_-}'(t)|\phi_{m_+,m_-}'(t)\rangle&=&\delta_{n_+,m_+}\delta_{n_-,m_-},\,\,\,\,\,\,\, n_-,n_+\in \mathbb{N}.
\end{eqnarray}
With the intention of determining the corresponding exact solution $\psi_{n_+,n_-}(x,y,t)$, we introduce the polar coordinates 
through the following canonical transformation  $x:r\cos\theta,\,\,y:r\sin\theta,\,
p_x:-i(\cos\theta \partial_r-\frac{\sin\theta}{r}\partial_\theta)$ and
$p_y:-i(\sin\theta \partial_r+\frac{\cos\theta}{r}\partial_\theta)$.
In terms of these representations the operators (\ref{lan11}) can be written as
\begin{eqnarray}
	{a'_\pm}^\dag&=&\frac{1}{2}e^{\mp i\theta}\left[\left(\frac{\kappa}{\rho}r-\rho\partial_r\right)\pm i\frac{\rho}{r}\partial_\theta\right],\\
	a'_\pm&=& \frac{1}{2}e^{\pm i\theta}\left[\left(\frac{\kappa}{\rho}r+\rho\partial_r\right)\mp i\frac{\rho}{r}\partial_\theta\right].
\end{eqnarray}
From the relation (\ref{lan13}) we construct the eigenfunctions for the invariant operator of the system
\cite{169}. One finds
\begin{eqnarray}\label{lan26}
	\psi_{n_+,n_-}(x,y,t)&=&U^\dag \psi'_{n_+,n_-}(x,y,t)\cr
	&=& (-)^n\frac{(\kappa)^{\frac{1+|\ell|}{2}}}{\rho^{1+|\ell|}\sqrt{\pi}}\sqrt{\frac{n!}{\Gamma(n+|\ell|+1)}} r^{|\ell|}\times\cr&& e^{\left(iM
		\frac{\dot{\rho}}{\rho}-\frac{\kappa}{\rho^2}\right)\frac{r^2}{2}}L_n^{|\ell|}\left(\frac{\kappa}{\rho^2}r^2\right)e^{i\ell\theta},
\end{eqnarray}
where $\ell=n_+-n_-$,\,\,\,$n=\min(n_+,n_-)=\frac{1}{2}(n_++n_--|\ell|)$, $\Gamma(u)$  and $L_n^{|\ell|}\left(u\right)$
are the Gamma function and  the generalised Laguerre polynomials of argument $u$, respectively.

To obtain the expectative values $E_{n_\pm},l_{n_\pm},\mathcal{E}_{n_\pm}$ of the operators $I(t), L_z, H(t)$ respectively, let us introduce a new
pair of raising and lowing helicity  operators define as
\begin{eqnarray}
	a_\pm&=&U^\dag a'_\pm U=\frac{1}{2\sqrt{\kappa}}\left[\left(M\dot{\rho}+i\frac{\kappa}{\rho}\right)(x\pm iy)-\rho(p_x\pm ip_y)\right],\\
	a_\pm^\dag&=&U^\dag {a'}_\pm^\dag U=\frac{1}{2\sqrt{\kappa}}\left[\left(M\dot{\rho}-i\frac{\kappa}{\rho}\right)(x\mp iy)-\rho(p_x\mp ip_y)\right].
\end{eqnarray}
Conversely
\begin{eqnarray}
	x&=&-\frac{i\rho}{2\sqrt{\kappa}}\left(a_--a_+^\dag+a_+-a_-^\dag\right),\\
	p_x&=&-\frac{iM\dot{\rho}}{2\sqrt{\kappa}}\left(a_--a_+^\dag+a_+
	-a_-^\dag\right)-\frac{\sqrt{\kappa}}{2\rho}\left(a_-+a_+^\dag+a_++a_-^\dag\right),\\
	y&=&\frac{\rho}{2\sqrt{\kappa}}\left(a_--a_+^\dag-a_++a_-^\dag\right),\\
	p_y&=&\frac{M\dot{\rho}}{2\sqrt{\kappa}}\left(a_--a_+^\dag-a_+
	+a_-^\dag\right)-i\frac{\sqrt{\kappa}}{2\rho}\left(a_-+a_+^\dag-a_+-a_-^\dag\right).
\end{eqnarray}
In particulary
\begin{eqnarray}
	x-iy&=&\frac{i\rho}{\sqrt{\kappa}}\left(a_+^\dag-a_-\right),\\
	p_x+ip_y&=&\frac{iM\dot{\rho}}{\sqrt{\kappa}}\left(a_-^\dag-a_+\right)
	-\frac{\sqrt{\kappa}}{\rho}\left(a_-^\dag+a_+\right), \\
	x+iy&=&\frac{i\rho}{\sqrt{\kappa}}\left(a_-^\dag-a_+\right),\\
	p_x-ip_y&=&\frac{iM\dot{\rho}}{\sqrt{\kappa}}\left(a_+^\dag-a_-\right)
	-\frac{\sqrt{\kappa}}{\rho}\left(a_+^\dag+a_-\right).
\end{eqnarray}
With these, we have
\begin{eqnarray}
	I(t)&=& \kappa\left(a_+^\dag a_++a_-^\dag a_-+\mathbf{I}\right),\\
	L_z&=&\left(a_-^\dag a_--a_+^\dag a_+\right),\\
	H(t)&=&\frac{1}{2\kappa}\left(M\dot{\rho}^2+\frac{\kappa^2}{M\rho^2}+M\Omega^2\rho^2\right)\left(a_+^\dag a_+ +a_-^\dag a_-+\mathbf{I}\right)
	\cr&&-\frac{\omega_c}{2}\left(a_-^\dag a_--a_+^\dag a_+\right)-\frac{q^2E^2}{2M\omega}\cr&& +\left(-\frac{M\dot{\rho}^2}{2\kappa}+i\frac{\dot{\rho}^2}{\rho}-\frac{\Omega^2M\rho^2}{2\kappa}
	+\frac{\kappa}{2M\rho^2}\right)a_-a_+\cr&&+
	\left(-\frac{M\dot{\rho}^2}{2\kappa}-i\frac{\dot{\rho}^2}{\rho}-\frac{\Omega^2M\rho^2}{2\kappa}
	+\frac{\kappa}{2M\rho^2}\right)a_-^\dag a_+^\dag,
\end{eqnarray}
and the angular momentum particulary satisfy the following relations
\begin{eqnarray}\label{lan24}
	[L_z,a_\pm]=\mp a_\pm,\,\,\,\,\ [L_z,a_\pm^\dag]=\pm a_\pm^\dag.
\end{eqnarray}
The expectative values of the above operators read  as
\begin{eqnarray}
	E_{n_\pm}&=&\langle \phi_{n_+,n_-}(t)|I(t)|\phi_{n_+,n_-}(t)\rangle=\kappa\left(n_++n_-+1\right), \\
	l_{n_\pm}&=&\langle \phi_{n_+,n_-}(t)|L_z|\phi_{n_+,n_-}(t)\rangle=n_--n_+,\\
	\mathcal{E}_{n_\pm}&=&\langle \phi_{n_+,n_-}(t)|H(t)|\phi_{n_+,n_-}(t)\rangle=
	\frac{1}{2\kappa}\left(M\dot{\rho}^2+\frac{\kappa^2}{M\rho^2}+M\Omega^2\rho^2\right)\left(n_+ +n_-+1\right)
	\cr&&-\frac{\omega_c}{2}\left(n_--n_+\right)-\frac{q^2E^2}{2M\omega},\label{lan16}
\end{eqnarray}
where the action of $a_\pm$ and $a_\pm^\dag$ on $ |\phi_{n_\pm}(t)\rangle$  finds expression in 
\begin{eqnarray}
	a_\pm^\dag|\phi_{n_\pm,n_\mp}(t)\rangle&=&\sqrt{n_\pm+1}|\phi_{n_\pm+1,n_\mp}(t)\rangle,\\
	a_\pm|\phi_{n_\pm,n_\mp}(t)\rangle&=& \sqrt{n_\pm}|\phi_{n_\pm-1,n_\mp}(t)\rangle,\\
	a_\pm^\dag a_\pm|\phi_{n_\pm,n_\mp}(t)\rangle&=& n_\pm|\phi_{n_\pm,n_\mp}(t)\rangle.
\end{eqnarray}
To determine the exact solution of the Schr\"{o}dinger equation (\ref{lan2}), 
we have to find the exact expression of the phase function in equation (\ref{lan14}) so that
\begin{eqnarray}
	\frac{d\gamma(t)}{dt}&=&\langle \phi_{n_+,n_-}(t)|i\frac{\partial }{\partial t}-H(t)|\phi_{n_+,n_-}(t)\rangle\cr
	&=& \langle \phi_{n_+,n_-}(t)|i\frac{\partial }{\partial t}|\phi_{n_+,n_-}(t)\rangle-
	\langle \phi_{n_+,n_-}(t)|H(t)|\phi_{n_+,n_-}(t)\rangle.\label{lan17}
\end{eqnarray}
Let us  evaluate the following expression
\begin{eqnarray}
	\langle \phi_{n_+,n_-}(t)|\frac{\partial }{\partial t}|\phi_{n_+,n_-}(t)\rangle&=&
	\frac{1}{\sqrt{n_+!n_-!}}
	\langle \phi_{n_+,n_-}(t)|\frac{\partial}{\partial t}\left[\left(a_+^\dag\right)^{n_+} \left(a_-^\dag\right)^{n_-}|\phi_{0,0}(t)\rangle\right]\cr
	&=&\langle \phi_{0,0}(t)|\frac{\partial}{\partial t}|\phi_{0,0}(t)\rangle+\frac{1}{\sqrt{n_+!n_-!}}\times\cr&&
	\langle \phi_{n_+,n_-}(t)|\frac{\partial}{\partial t}\left[\left(a_+^\dag\right)^{n_+} \left(a_-^\dag\right)^{n_-}\right]|\phi_{0,0}(t)\rangle.
\end{eqnarray}
On  the one hand, we have
\begin{eqnarray}
	\langle \phi_{0,0}(t)|\left[\frac{\partial}{\partial t}|\phi_{0,0}(t)\rangle\right]&=&
	\int dxdy\phi_{0,0}^*(x,y)\frac{\partial \phi_{0,0}(x,y)}{\partial t}\cr
	&=&\int dxdy|\phi_{0,0}(t)|^2\cr&&\times\left[\left(i\dot{M}\frac{\dot{\rho}}{\rho}+iM\frac{\ddot{\rho}}{\rho}-iM\frac{\dot{\rho}^2}{\rho^2}
	+2\kappa\frac{\dot{\rho}}{\rho^3}\right)\frac{x^2+y^2}{2}-\frac{\dot{\rho}}{\rho} \right].
\end{eqnarray}
By the Stokes theorem such as  
\begin{eqnarray}
	\int dxdy|\phi_{0,0}(t)|^2(x^2+y^2)=\frac{\rho^2}{\kappa}\int dxdy|\phi_{0,0}(x,y)|^2=\frac{\rho^2}{\kappa},
\end{eqnarray}
and some straightforward computations we obtain
\begin{equation}
	\langle \phi_{0,0}(t)|\left[\frac{\partial}{\partial t}|\phi_{0,0}(t)\rangle\right]   = \frac{iM}{2\kappa}\left((\ddot{\rho}\rho+\dot{\rho}\rho-\dot{\rho}^2\right).
\end{equation}
On the other hand,
\begin{eqnarray}
	\frac{1}{\sqrt{n_+!n_-!}}\langle \psi_{n_+,n_-}(t)|\frac{\partial}{\partial t}\left[\left(a_+^\dag\right)^{n_+}
	\left(a_-^\dag\right)^{n_-}\right]|\psi_{0,0}(t)\rangle&=&
	\frac{iM}{2\kappa}\left(\ddot{\rho}\rho+\frac{\dot{M}}{M}\rho\dot{\rho}-\dot{\rho}^2\right)\cr&&\times(n_++n_-),
\end{eqnarray}
where the expressions of $\frac{\partial a_+^\dag}{\partial t}$ and
$\frac{\partial a_-^\dag}{\partial t}$ in terms of $a_\pm$ and $a_\pm^\dag$ are given by
\begin{eqnarray}
	\frac{\partial a_+^\dag}{\partial t}&=&\frac{1}{2\sqrt{\kappa}}\left[\left(\dot{M}\dot{\rho}+
	M\ddot{\rho}+ik\frac{\dot{\rho}}{\rho^2}\right)(x-iy)-\dot{\rho}(p_x-ip_y)\right]\cr
	&=&\frac{iM}{2\kappa}\left(\ddot{\rho}\rho+\frac{\dot{M}}{M}\rho\dot{\rho}-\dot{\rho}^2\right)a_+^\dag+\left[\frac{\dot{\rho}}{\rho}
	-\frac{iM}{2\kappa}\left(\rho \ddot{\rho}+\frac{\dot{M}}{M}\rho-\dot{\rho}^2\right)\right]a_-,\\
	\frac{\partial a_-^\dag}{\partial t}&=&\frac{1}{2\sqrt{\kappa}}\left[\left(\dot{M}\dot{\rho}+
	M\ddot{\rho}+ik\frac{\dot{\rho}}{\rho^2}\right)(x+iy)-\dot{\rho}(p_x+ip_y)\right]\cr
	&=& \frac{iM}{2\kappa}\left(\ddot{\rho}\rho+\frac{\dot{M}}{M}\rho\dot{\rho}-\dot{\rho}^2\right)a_-^\dag+\left[\frac{\dot{\rho}}{\rho}
	-\frac{iM}{2\kappa}\left(\rho \ddot{\rho}+\frac{\dot{M}}{M}\rho-\dot{\rho}^2\right)\right]a_+.
\end{eqnarray}
We obtain
\begin{eqnarray}
	\langle \phi_{n_+,n_-}(t)|\frac{\partial}{\partial t}|\phi_{n_+,n_-}(t)\rangle&=& \frac{iM}{2\kappa}\left(\ddot{\rho}\rho+
	\frac{\dot{M}}{M}\rho\dot{\rho}-\dot{\rho}^2\right)(n_++n_- +1)\cr
	&=&\frac{iM}{2\kappa}\left(\frac{\kappa^2}{M^2\rho^2}-\Omega^2\rho^2-\dot{\rho}^2\right)
	\cr&&\times(n_++n_-+1).\label{lan15}
\end{eqnarray}
Finally, taking into account (\ref{lan15}), (\ref{lan16}) and (\ref{lan17}), we find  that the phase function  is given by
\begin{eqnarray}
	\gamma (t)&=&-\frac{\kappa}{2}\left(n_++n_-+1\right)\int_0^t\frac{dt'}{M(t')\rho^2(t')}
	\cr&&+\frac{1}{2}(n_--n_+)\int_0^tdt' \omega_c(t')+\frac{q^2}{2}
	\int_0^t \frac{E^2(t')}{M(t')\omega(t')}dt'.
\end{eqnarray}
Our result for $\gamma(t)$ differs slightly  from the one calculated in \cite{104}
as the term $\frac{q^2}{2}  \int_0^t \frac{E^2(t')}{M(t')\omega(t')} $ doesn't appear in their result. This is  due to their  choice of  gauge.
Concerning the results obtained in Refs \cite{105,106,107}, the comparison shows  that  the difference is more large.
Ideed, in  addition to time dependent electric field contribution $\frac{q^2}{2}  \int_0^t \frac{E^2(t')}{M(t')\omega(t')} $, 
the angular  momentum  contribution $\frac{1}{2}(n_--n_+)\int_0^tdt' \omega_c(t')$ doesn't also appear. This is due to the analytical 
method used. Finally this result for $\gamma(t)$  is reduced to our recently result \cite{194} in absence of electromagnetism field.\\
The solution of the  Schr\"{o}dinger equation is   given by
\begin{eqnarray}
	\psi_{n_+,n_-}(x,y,t)&=&(-)^n\frac{(\kappa)^{\frac{1+|\ell|}{2}}}{\rho^{1+|\ell|}\sqrt{\pi}}\sqrt{\frac{n!}{\Gamma(n+|\ell|+1)}} r^{|\ell|}\times\cr&& e^{\left(iM
		\frac{\dot{\rho}}{\rho}-\frac{\kappa}{\rho^2}\right)\frac{r^2}{2}}L_n^{|\ell|}\left(\frac{\kappa}{\rho^2}r^2\right)e^{il\theta} 
	e^{i\gamma(t)}. 
\end{eqnarray}

Moreover, this work arouses in our mind the following question what happens in  the case  of the three dimensional confinement 
of the  charged particle? In this way, the time-dependent Hamiltonian  of the system in this  new context reads  as
\begin{eqnarray}
	\mathcal{H}(t)&=& \frac{1}{2M(t)}\left[\left(p_1+\frac{q}{2}B(t)x_2\right)^2+\left(p_2-\frac{q}{2}B(t)x_1\right)^2+p_3^2\right]+\cr&&
	\frac{1}{2} M(t) \omega(t)(x_1^2+x_2^2+x_3^2) -q[E_1(t)x_1+E_2(t)x_2].
\end{eqnarray}
Performing the same transformations (\ref{lan18}),(\ref{lan19}) and for being in the condition where  the magnetic field is weak 
($\Omega(t)\simeq \omega (t)$) for
technical reason, the Hamiltonian of the  system becomes
\begin{eqnarray}
	\mathcal{H}(t)&=&\frac{1}{2M(t)}(p_x^2+p_y^2+p_z^2)+\frac{1}{2}\Omega^2(t)M(t)(x^2+y^2+z^2)\cr&&-
	\frac{\omega_c (t)}{2}L_z-\frac{q^2E^2(t)}{2M(t)\Omega(t)}.
\end{eqnarray}
By proceeding as in the previous  section, the 3D invariant operator is written in the form
\begin{eqnarray}\label{lan21}
	\mathcal{I}(t)=\frac{1}{2}\left[\frac{\kappa^2}{\rho^2}(x^2+y^2+z^2)+\left(\rho p_x-M\dot{\rho}x\right)^2+
	\left(\rho p_y-M\dot{\rho}y\right)^2+ \left(\rho p_z-M\dot{\rho}z\right)^2\right],
\end{eqnarray}
with $\rho$ satisfying the  same equation (\ref{lan8}).\\
Applying the  unitary transformation (\ref{lan20}), one reduces the invariant operator (\ref{lan21}) as follows
\begin{eqnarray}
	\mathcal{I}'(t)= W^\dag \mathcal{I}(t) W= \frac{1}{2}\left[\rho^2(p_x^2+p_y^2+p_z^2)+\frac{\kappa^2}{\rho^2}(x^2+y^2+z^2)\right],
\end{eqnarray}
with
\begin{equation}
	W= \exp[-i\frac{M\dot{\rho}}{2\rho}(x^2+y^2+z^2)].
\end{equation} 
Instead of choosing to diagonalize  the invariant operator 
through analytical  method  proceeding by the cylinder parametrization $(r, \theta, z)$ of the degrees of freedom of the system,
we rather  proceed by algebraic method to solve  the eigenvalue equation for the invariant operator.  
The invariant operator and the angular momentum read
\begin{eqnarray}
	\mathcal{I}'(t)=\frac{1}{2}\rho^2p_z^2+ \kappa\left({a'_+}^\dag a'_++{a'_-}^\dag a'_-+\mathbf{I}\right),\,\,\,
	L_z={a'_-}^\dag a'_--{a'_+}^\dag a'_+.
\end{eqnarray}
The eigenvalues of both operators are given by
\begin{eqnarray}
	\mathcal{I}'(t)|\phi'_{n_\pm,p}(t)\rangle &=& \alpha_{n_\pm,p}|\phi'_{n_\pm,p}(t)\rangle,\,
	\alpha_{n_\pm,p}=\frac{1}{2}\rho^2p^2+\kappa\left(n_++n_-+1\right),\label{lan22}\\
	L_z|\phi'_{n_\pm,p}(t)\rangle &=& l_{n_\pm,p}|\phi'_{n_\pm,p}(t)\rangle,\,\,\,\,\,\,\,\, l_{n_\pm,p}=n_--n_+,
\end{eqnarray}
with $-\infty<p<\infty$.\\
One can remark  that, the eigenvalues of the invariant operator (\ref{lan22})  are time dependent while, in fact, they should be time independent
according to the Lewis-Riesenfeld theory \cite{47,47'}. As Maamache \textit{et al} \cite{105} remarked that 
result of Ferreira \textit{et al}  \cite{107} contradicts  
the Lewis-Riesenfeld method, we deduce that, the 
$3$D of the previous model (\ref{lan23}) can be a limit of the  Lewis-Riesenfeld method.

\subsection{Standard canonical coherent states}
Given any complex values $z_\pm \in \mathbb{C}$  (related to the operator $a_\pm,\,a_\pm^\dag$ ), such as
\begin{eqnarray}
	z_\pm&=&\frac{1}{2\sqrt{\kappa}}\left[\left(M\dot{\rho}+i\frac{\kappa}{\rho}\right)(x\pm iy)-\rho(p_x\pm ip_y)\right],\\
	z_\pm^*&=&\frac{1}{2\sqrt{\kappa}}\left[\left(M\dot{\rho}-i\frac{\kappa}{\rho}\right)(x\mp iy)-\rho(p_x\mp ip_y)\right].
\end{eqnarray}%
The coresponding  helicity (holomorphic) normalised  coherent states are defined by
\begin{eqnarray}
	|\psi_{z_-,z_+}\rangle&=& |\psi_{z_-}\rangle\otimes|\psi_{z_+}\rangle,\cr
	&=&e^{z_+a_+^\dag- z_+^* a_+}e^{z_-a_-^\dag- z_-^* a_-}|\psi_{00}\rangle,\cr
	&=& e^{-\frac{1}{2}|z_-|^2-\frac{1}{2}|z_+|^2}\sum_{n_-,n_+=0}^\infty \frac{(z_-)^{n_-}(z_+)^{n_+}}{\sqrt{n_-!n_+!}}|\psi_{n_+,n_-}\rangle,
\end{eqnarray}
In coordinate representation for $\ell>0$, the wavefunction for the  coherent states  is given by
\begin{eqnarray}
	\psi_{z_\pm}(x,y,t)&=&\sum_{n_-,n_+=0}^\infty(-)^{n_-}\frac{(\kappa)^{\frac{1+n_+-n_-}{2}}}{\rho^{1+n_+-n_-}\sqrt{\pi}}e^{-\frac{1}{2}|z_-|^2-\frac{1}{2}|z_+|^2}
	\sqrt{\frac{n_-!}{\Gamma(n_++1)}} r^{n_+-n_-}\times \cr&&  \frac{(z_-)^{n_-}(z_+)^{n_+}}{\sqrt{n_-!n_+!}}
	e^{\left(iM\frac{\dot{\rho}}{\rho}-\frac{\kappa}{\rho^2}\right)\frac{r^2}{2}}L_{n_-}^{n_+-n_-}
	\left(\frac{\kappa}{\rho^2}r^2\right)e^{in_+\theta}  e^{i\gamma(t)}.
\end{eqnarray}
The non-orthogonality of these states  are given as follows
\begin{eqnarray}
	\langle \psi_{\alpha_-,\alpha_+} |\psi_{z_-,z_+}\rangle=e^{-\frac{1}{2}|\alpha_--z_-|^2}e^{-\frac{1}{2}|\alpha_+-z_+|^2}.
\end{eqnarray}
In terms of these generating vectors, the resolution of the identity is expressed as
\begin{eqnarray}
	\int \frac{dz_+ d z_+^* dz_-d z_-^*}{\pi^2}|\psi_{z_+,z_-}\rangle\langle\psi_{z_+,z_-}|= \mathbb{I}=
	\sum_{n_+,n_-=0}^\infty|\psi_{n_+,n_-}\rangle\langle\psi_{n_+,n_-}|,
\end{eqnarray}
with the basic matrix elements for the change of bases given by
\begin{eqnarray}
	\langle \psi_{n_+,n_-}|\psi_{z_-,z_+}\rangle &=& e^{-\frac{1}{2}|z_-|^2-\frac{1}{2}|z_+|^2}\sum_{n_-,n_+=0}^\infty \frac{(z_-)^{n_-}(z_+)^{n_+}}{\sqrt{n_-!n_+!}}\\
	\langle \psi_{z_+,z_-}|\psi_{n_-,n_+}\rangle &=& e^{-\frac{1}{2}|z_-|^2-\frac{1}{2}|z_+|^2}\sum_{n_-,n_+=0}^\infty \frac{( z_-^*)^{n_-}( z_+^*)^{n_+}}
	{\sqrt{n_-!n_+!}}.
\end{eqnarray}
From a straightforward analysis using (\ref{lan24}) , the action of the $SO(2)=U(1)$ 
rotation generator $L_z$ on the coherent states is 
\begin{equation}
	e^{i\alpha L_z}|\psi_{z_\pm}\rangle= |e^{\pm i\alpha }\psi_{z_\pm}\rangle,
\end{equation}
where $\alpha\in \mathbb{R}$.

The  actions of the operators $a_\pm$ on  $|\psi_{z_-,z_+}\rangle$ satisfy the following properties
\begin{eqnarray}
	a_\pm|\psi_{z_-,z_+}\rangle=z_\pm|\psi_{z_-,z_+}\rangle,\,\,\, \langle \psi_{z_-,z_+}| a_\pm= \langle \psi_{z_-,z_+}|\partial_{  z_\pm^*},
\end{eqnarray}
and for the creator operators we have 
\begin{eqnarray}
	a_\pm^\dag|\psi_{z_-,z_+}\rangle=\partial_{z_\pm}|\psi_{z_-,z_+}\rangle
	,\,\,\, \langle \psi_{z_-,z_+}| a_\pm^\dag= \langle \psi_{z_-,z_+}|  z_\pm^*
\end{eqnarray}
The  actions of the operators $a_\pm$ on  $|\psi_{z_-,z_+}\rangle$ satisfy the following properties
\begin{eqnarray}
	l(z_+,z_+, z_+^*, z_-^*)&=& |z_-|^2-|z_+|^2,\\
	h(z_+,z_+, z_+^*, z_-^*)&=&\frac{\langle \psi_{z_-,z_+}|H(t)|\psi_{z_-,z_+}\rangle}{\langle \psi_{n_-,n_+}|\psi_{z_-,z_+}\rangle},\\
	&=&\frac{1}{2\kappa}\left(M\dot{\rho}^2+\frac{\kappa^2}{M\rho^2}+M\Omega^2\rho^2\right)\left(|z_+|^2 +|z_-|^2+1\right)
	\cr&&-\frac{\omega_c}{2}\left(|z_-|^2-|z_+|^2\right)-\frac{q^2E^2}{2M\omega}\cr&& +\left(-\frac{M\dot{\rho}^2}{2\kappa}+i\frac{\dot{\rho}^2}{\rho}-\frac{\Omega^2M\rho^2}{2\kappa}
	+\frac{\kappa}{2M\rho^2}\right)z_-z_+\cr&&+
	\left(-\frac{M\dot{\rho}^2}{2\kappa}-i\frac{\dot{\rho}^2}{\rho}-\frac{\Omega^2M\rho^2}{2\kappa}
	+\frac{\kappa}{2M\rho^2}\right) z_-^* z_+^*.
\end{eqnarray}
The time evolution of the system reads as 
\begin{eqnarray}
	|\psi_{z_-,z_+}(\tau)\rangle&=&e^{-iH\tau}|\psi_{z_-,z_+}\rangle,\cr
	&=&e^{-\frac{1}{2}|z_-|^2-\frac{1}{2}|z_+|^2}\sum_{n_-,n_+=0}^\infty \frac{(z_-)^{n_-}(z_+)^{n_+}}{\sqrt{n_-!n_+!}}
	e^{-i\tau E_{n_+,n_-}}|\psi_{n_+,n_-}\rangle,\cr
	&=&e^{-\frac{1}{2}|z_-|^2-\frac{1}{2}|z_+|^2}e^{-i(T_1-\lambda)\tau}\cr&&\times
	\sum_{n_-,n_+=0}^\infty\frac{[a_-^\dag e^{-i(T_1-T_2)\tau}z_-]^{n_-}}{n_-!}
	\frac{[a_+^\dag e^{-i(T_1+T_2)\tau}z_+]^{n_+}}{n_+!}|\psi_{00}\rangle\cr
	&=& e^{-i(T_1-\lambda)\tau}| e^{-i(T_1-T_2)\tau}e^{-i(T_1+T_2)\tau}\psi_{z_+z_-}\rangle,
\end{eqnarray}
where $T_1=\frac{1}{2\kappa}\left(M\dot{\rho}^2+\frac{\kappa^2}{M\rho^2}+M\Omega^2\rho^2\right)$,\,\,\,\,
$T_2= \frac{\omega_c}{2}$ and $\lambda=\frac{q^2E^2}{2M\omega}$.

The Heisenberg uncertainty relations for the simultaneous measurement of the observables $A$ and $B$ in 
the state $\{|\psi_{n_+,n_-}\rangle\}$ has to obey the inequality
\begin{eqnarray} \label{v3}
	\Delta A&=&\sqrt{\langle\psi_{n_+,n_-}|A^2|\psi_{n_+,n_-}\rangle-\langle\psi_{n_+,n_-}|A|\psi_{n_+,n_-}\rangle^2},\\
	\Delta B&=&\sqrt{\langle\psi_{n_+,n_-}|B^2|\psi_{n_+,n_-}\rangle-\langle\psi_{n_+,n_-}|B|\psi_{n_+,n_-}\rangle^2}.
\end{eqnarray}
As far as  our context is concerned,
the standard expectation values of operators $x,y,p_x,p_y$  are evaluating  as follows
the Heisenberg uncertainty relations   can be inferred 
\begin{eqnarray}
	\Delta x\Delta p_x=\Delta y \Delta p_y=\frac{1}{2}\left(2n+|\ell|+1\right)\sqrt{1+\frac{M^2\dot{\rho}^2\rho^2}{\kappa^2}}\geq \frac{1}{2}.
\end{eqnarray}
We can conclude that from this  relation the helicity Fock states do not minimize the Heisenberg incertainty. But for $n=0$ and $\ell=0$ 
we have the following relation
\begin{eqnarray}
	\Delta x\Delta p_x=\Delta y \Delta p_y=\frac{1}{2}\sqrt{1+\frac{M^2\dot{\rho}^2\rho^2}{\kappa^2}}
\end{eqnarray}
which minimizes the Heisenberg incertainty. 

The  particle distribution is given by the Poissonian function
\begin{eqnarray}
	P_{z_\pm}(n_+,n_-)&=&|\langle \psi_{n_\pm}|\psi_{z_\pm}\rangle|^2\cr
	&=&e^{-|z_-|^2-|z_+|^2} \frac{|z_-|^{2n_-}|z_+|^{2n_+}}{n_-!n_+!}.
\end{eqnarray}

\subsection{Nonlinear canonical coherent states}
We introduce two nonlinear Fock algebras of the type constructed by Jannussis \textit{et al} \cite{184}.
These algebras are factorised with respect to helicity and are defined by two real functions 
$f_\pm(N_\pm)\neq 0$ and the generators $A_\pm$ , $A_\pm^\dag$ such that
\begin{eqnarray}
	A_\pm&=&a_\pm f_\pm(N_\pm),\,\,\,\,A_\pm^\dag= f_\pm(N_\pm)a_\pm^\dag,\\
	{ [A_\pm,A_\pm^\dag]}&=&(N_\pm+1)f_\pm^2(N_\pm+1)-(N_\pm)f_\pm^2(N_\pm),
\end{eqnarray}
where $N_\pm=a_\pm^\dag a_\pm$.
The $f_\pm$-coherent states $|\psi_{ \alpha_\pm,f_\pm}\rangle$ are defined by the following equation
\begin{equation}\label{nonl1}
	A_\pm|\psi_{ \alpha_\pm,f_\pm}\rangle=\alpha_\pm|\psi_{ \alpha_\pm,f_\pm}\rangle.
\end{equation}
Referring to the previous chapter   in the case of the construction of one mode nonlinear coherent states, we deduce
\begin{eqnarray}
	|\psi_{ \alpha_\pm,f_\pm}\rangle=\mathcal{N}_{\alpha_\pm,f_\pm} \sum_{n_+=0}^\infty \sum_{n_-=0}^\infty\frac{(\alpha_+)^{n_+}(\alpha_-)^{n_-}}
	{\sqrt{n_-!n_+!}[f_-(n_-)]![f_+(n_+)]!}|\psi_{n_\pm}\rangle
\end{eqnarray}
where 
\begin{equation}
	\mathcal{N}_{\alpha_\pm,f_\pm} =\sum_{n_+=0}^\infty \sum_{n_-=0}^\infty \left(|\alpha_+|^{2n_+}|\alpha_-|^{2n_-}n_-!n_+! \left([f_-(n_-)]![f_+(n_+)]!\right)^{-2}\right)^{-\frac{1}
		{2}}.
\end{equation}
In the case $f_1=f_2=1$, we recover the above  two-mode coherent states, namely,
\begin{eqnarray}
	|\psi_{ \alpha_\pm,1}\rangle= e^{-(|\alpha_+|^2+|\alpha_-|^2)/2}\sum_{n_+=0}^\infty \sum_{n_-=0}^\infty
	\frac{(\alpha_+)^{n_+}(\alpha_-)^{n_-}}
	{\sqrt{n_-!n_+!}}|\psi_{n_\pm}\rangle.
\end{eqnarray}
As we  can  see, the states  (\ref{nonl1})  are not orthogonal
\begin{eqnarray}
	\langle \psi_{ \sigma_\pm,g_\pm} |\psi_{ \alpha_\pm,f_\pm}\rangle&=&\mathcal{N}_{\alpha_\pm,f_\pm} 
	\mathcal{N}_{\sigma_\pm,g_\pm} \sum_{n_-,n_+=0}^\infty\sum_{m_-,m_+=0}^\infty \delta_{n_\pm,m_\pm}\cr&&\times
	\frac{(\alpha_+)^{n_+}(\alpha_-)^{n_-}(\sigma_+^*)^{m_+}(\sigma_-^*)^{m_-}} {\sqrt{n_-!n_+!m_-!m_+!}
		[f_-(n_-)]![f_+(n_+)]![g_-(m_-)]![g_+(m_+)]!}\cr
	&=& \mathcal{N}_{\alpha_\pm,f_\pm} 
	\mathcal{N}_{\sigma_\pm,g_\pm}\cr&&\times\sum_{n_-,n_+=0}^\infty
	\frac{(\alpha_+\sigma_+^*)^{n_+}(\alpha_-\sigma_-^*)^{n_-}} {n_-!n_+!
		[f_-(n_-)]![f_+(n_+)]![g_-(n_-)]![g_+(n_+)]!}.
\end{eqnarray}
The resolution of  identity  associated to these nonlinear coherent states  satisfy  
\begin{eqnarray}\label{nor6}
	\int_{\mathbb{C}}\frac{ \mathcal{N}_{\alpha_\pm,f_\pm}}{\pi^2}|\alpha_\pm,f_\pm\rangle\langle f_\pm,\alpha_\pm|\Omega(|\alpha_\pm|^2)d^2\alpha_\pm=\mathbb{I}.
\end{eqnarray}
This completeness holds for the defined  position weight function  $\Omega (t)$
such as 
\begin{equation}\label{nor9}
	\int_0^\infty t^{n_++n_-} \Omega^2 (t)d^2t=n_-!n_+![f_-(n_-)!]^2[f_+(n_+)!]^2.
\end{equation}

Since the state $|\alpha_\pm,f_\pm\rangle$ is given as series of Fock  helicity   states, we can easily write the wave function
of these states in different representations explicitly.
In coordinate representation, the wave function is
\begin{eqnarray}
	\psi_{\alpha_\pm,f_\pm}^{\ell,n}(x,y,t)&=&(-)^n\frac{(\kappa)^{\frac{1+|\ell|}{2}}}{\rho^{1+|\ell|}
		\sqrt{\pi}}\mathcal{N}_\pm\sqrt{\frac{n!}{\Gamma(n+|\ell|+1)}} r^{|\ell|}\times e^{\left(iM
		\frac{\dot{\rho}}{\rho}-\frac{\kappa}{\rho^2}\right)\frac{r^2}{2}}\sum_{n_+=0}^\infty \sum_{n_-=0}^\infty\cr&&\times\frac{(\alpha_+)^{n_+}(\alpha_-)^{n_-}}
	{\sqrt{n_-!n_+!}[f_-(n_-)]![f_+(n_+)]!}
	L_n^{|\ell|}\left(\frac{\kappa}{\rho^2}r^2\right)e^{i\ell\theta} 
	e^{i\gamma(t)}. 
\end{eqnarray}
For the Bargmann representation (the usual coherent states), the wave function $\langle \psi_{z_+,z_-}|\psi_{\alpha_\pm,f_\pm} \rangle$,
where we use the bases $|\psi_{z_\pm}\rangle $ $(z_\pm \in \mathbb{C})$ takes the form
\begin{eqnarray}
	\psi_{\alpha_\pm,f_\pm}(z_\pm)=\mathcal{N}_\pm e^{-\frac{1}{2}|z_-|^2-\frac{1}{2}|z_+|^2}\sum_{n_+=0}^\infty \sum_{n_-=0}^\infty
	\frac{(z_+^*\alpha_+)^{n_+} (z_-^*\alpha_-)^{n_-}}{n_-!n_+![f_-(n_-)]![f_+(n_+)]!}
\end{eqnarray}
The particle distribution given  by the Poissonian function in the case  $f_\pm$-coherent states becomes
\begin{eqnarray}
	P_{\alpha_\pm,f_\pm}=\left(\sum_{j_+=0}^\infty \sum_{j_-=0}^\infty\frac{|\alpha_+|^{2j_+}|\alpha_-|^{2j_-}}{j_-!j_+![f_+(j_+)!]^2[f_-(j_-)!]^2}\right)^{-1}
	\frac{|\alpha_+|^{2n_+}|\alpha_-|^{2n_-}}{n_+!n_-![f_-(n_-)!]^2[f_+(n_+)!]^2}.
\end{eqnarray}
\subsection{Photon added coherent states}
The two-mode linear  Photon added coherent states of the one mode we defined in the above section are defined as follows
\begin{equation}
	|\psi_{ \alpha_\pm,m_\pm}\rangle=\mathcal{N}_{\alpha_\pm,m_\pm}(a_-^\dag)^{m_-}(a_+^\dag)^{m_+}|\psi_{ \alpha_\pm}\rangle,
\end{equation}
where $m_\pm$ are  positive integers being the numbers of added quanta (or added photons) and \\ 
$\mathcal{N}_{\alpha_\pm,m_\pm}=\left[\langle \psi_{ \alpha_\pm}|a_-^{m_-}a_+^{m_+}( a_-^\dag)^{m_-}(a_+^\dag)^{m_+}|\psi_{\alpha_\pm}\rangle\right]^{-\frac{1}{2}}$.
These states in  terms of Fock states can be written as
\begin{eqnarray}
	|\psi_{ \alpha_\pm,m_\pm}\rangle&=& \mathcal{N}_{\alpha_\pm,m_\pm}(|\alpha_\pm|)\sum_{n_+,n_-=0}^\infty
	\frac{(\alpha_-)^{n_-}(\alpha_+)^{n_+}\sqrt{(n_-+m_-)!(n_++m_+)!}}{n_-!n_+!}\cr &&\times|\psi_{n_-+m_-,n_++m_+}\rangle.
\end{eqnarray}
where $ \mathcal{N}_{\alpha_\pm,m_\pm}(|\alpha_\pm|)=\mathcal{N}_{\alpha_\pm,m_\pm}e^{-\frac{|\alpha_-|^2}{2}-\frac{|\alpha_+|^2}{2}}$.
The non-orthogonality of these states reads  as
\begin{eqnarray}
	\langle \psi_{ \sigma_\pm,m_\pm} |\psi_{ \alpha_\pm,m_\pm}\rangle&=&\mathcal{N}_{\alpha_\pm,m_\pm}(|\alpha_\pm|)
	\mathcal{N}_{\alpha_\pm,m_\pm}(|\sigma_\pm|)\sum_{n_+,n_-=0}^\infty\cr&&\times
	\frac{(\alpha_-\sigma_-^*)^{n_-}(\alpha_+\sigma_+^*)^{n_+}(n_-+m_-)!(n_++m_+)!}{[n_-!]^2[n_+!]^2}
\end{eqnarray}

The two-modes photon coherent states  are eigenstates of the annihalation operators $a_+,\,\,\,a_-$ given by the  following equation
\begin{equation}
	a_+a_-  |\psi_{ \alpha_\pm,m_\pm}\rangle=\alpha_\pm |\psi_{ \alpha_\pm,m_\pm}\rangle.
\end{equation}
Multiplying both sides by $(a_-^\dag)^{m_-}(a_+^\dag)^{m_+}$
\begin{eqnarray}
	(a_-^\dag)^{m_-}(a_+^\dag)^{m_+}a_+a_- |\psi_{ \alpha_\pm,m_\pm}\rangle=\alpha_\pm 
	(a_-^\dag)^{m_-}(a_+^\dag)^{m_+}|\psi_{ \alpha_\pm,m_\pm}\rangle,
\end{eqnarray}
which, by making use of the commutation relations $[a_+, (a_+^\dag)^{m_+}] = m_+ (a^\dag)^{m_+-1}$,\,\,\,\,
$[a_-, (a_-^\dag)^{m_-}] = m_- (a^\dag)^{m_--1}$,
and the identity
\begin{eqnarray}
	(a_+^\dag)^{m_+}a_+&=&a_+ (a_+^\dag)^{m_+}-m_+(a_+^\dag)^{m_+-1},\\
	(a_-^\dag)^{m_-}a_-&=&a_- (a_-^\dag)^{m_-}-m_-(a_-^\dag)^{m_--1},
\end{eqnarray}
it leads to
\begin{equation}\label{lan25}
	\left(1-\frac{m_+}{a_+^\dag a_++1}\right)\left(1-\frac{m_-}{a_-^\dag a_-+1}\right)a_+a_- 
	|\psi_{ \alpha_\pm,m_\pm}\rangle=\alpha_\pm |\psi_{ \alpha_\pm,m_\pm}\rangle.
\end{equation}
From (\ref{lan25}) we can see that the  two-mode photon-added  coherent states arise  two-mode
class of nonlinear coherent with the corresponding nonlinear function 
\begin{equation}
	f(N_+,N_-)= \left(1-\frac{m_+}{a_+^\dag a_++1}\right)\left(1-\frac{m_-}{a_-^\dag a_-+1}\right),
\end{equation}
with $N_\pm=a_\pm^\dag a_\pm.$
Another aspect of these states are  revealed by considering their photon number distributions $P_{\alpha_\pm}(n_\pm,m_\pm)$
which is the probability of finding an oscillator in the state $|\psi_{ n_\pm,m_\pm}\rangle$
defined as follows
\begin{eqnarray}
	P_{\alpha_\pm}(n_\pm,m_\pm)&=&|\langle \psi_{ n_++m_+,n_-+m_-}|\psi_{ \alpha_\pm,m_\pm}\rangle|^2,\cr
	&=&\mathcal{N}_{\pm,m_\pm}^2 
	\frac{|\alpha_-|^{2n_-}|\alpha_+|^{2n_+}(n_-+m_-)!(n_++m_+)!}{[n_-!]^2[n_+!]^2}.
\end{eqnarray}

\section{SU(2) symmetry and its coherent states for the model}
\subsection{SU(2) symmetry of the model}

As matter of fact, one  observed   the degeneracy of energy  $E_{n_\pm}$ or  $E_{n_x,n_y}$  of the invariant operator at each levels in term of a fixed value $n=n_x+n_y=n_++n_-$ 
except for the ground state. We have as many states as there are partitions of the natural number $n$ in two natural
numbers $n_1,n_2$ or $n_+,n_-$ namely $(n + 1)$-dimension states. Therefore, this fact cannot be just a mere numerical coincidence, the must  exist a
solid explanation for this fact. Indeed even though, the system possesses a  $SO(2)=U(1)$ symmetry, this degeneracy cannot be related to it, since
the angular-momentum operator does not map states belonging to a same level into one another.

The first thought that comes to one's mind is that of a symmetry. As a matter of fact, the  system possesses a global symmetry
larger than the $SO(2)=U(1)$ rotational symmetry considered which explains the degeneracies of the energy spectrum $E(n_x,n_y)=E(n_+,n_-)=E(n)=\kappa(n+1)$.
This larger dynamical symmetry is  the $SU(2)$ symmetry or   the $SO(3)$ symmetry which are identical algebras but 
not as groups since $SO(3)=SU(2)/\mathbb{Z}_2$  as a quotient of groups. This $SU(2)$ symmetry is related to the possibility
of performing arbitrary $SU(2)$ rotations among the creation (or the annihila-
tion) operators $a_i$ (or $a_i$) ($i=1,2$), or $a_\pm^\dag$  (or $ a_\pm)$. The generators of this  algebra 
\begin{equation}
	J_k=\frac{\sigma_k}{2}
	,\,\,\,\, k=1,2,3
\end{equation}
where $\sigma_k$ are the usual Pauli matrices such that
\begin{eqnarray}
	\sigma_1= \begin{pmatrix}
		0&1 \\ 1&0
	\end{pmatrix},\,\,\,
	\sigma_2= \begin{pmatrix}
		0&-i \\ i&0
	\end{pmatrix}, \,\,\,
	\sigma_3= \begin{pmatrix}
		1&0 \\ 0&-1
	\end{pmatrix}.
\end{eqnarray}

These operators generate the element $SU(2)$  Lie group in the same way  that $L_z$ generate $U(1)$ group
\begin{equation}
	\mathcal{R}(\theta)=\exp\left(-i\theta_k J_k\right),\,\,\,\, \theta_k\in \mathbb{R}.
\end{equation}
The commutation  relation satisfied by the generators $J_k$  is written
\begin{eqnarray}
	[J_i,J_j]=i\varepsilon_{ijk} J_k,\,\,\,\,\,\, i,j,k=1,2,3.
\end{eqnarray}

Let us now see how we may introduce, based on the helicity Fock algebra generators $a_\pm$ and $a_\pm^\dag$
operators  which described the degeneracies of the  energy level of the invariant operator $I(t)$. Indeed, we introduce the raising $J_+$
and lowering $J_-$ operators defined as follows
\begin{eqnarray}
	J_\pm=J_1\pm i J_2.
\end{eqnarray}
Therefore, we have
\begin{eqnarray}
	J_+=a_+^\dag a_-,\,\,\,J_-=a_-^\dag a_+,\,\,\,\, J_3=\frac{1}{2}(a_+^\dag a_+-a_-^\dag a_-)=\frac{L_z}{2}.
\end{eqnarray}
It takes but only a little calculation to obtain
\begin{eqnarray}
	[J_+,J_-]=2J_3,\,\,\,[J_3,J_\pm]=\pm J_\pm.
\end{eqnarray}
The action of the operators $J_\pm$ on the states $|\psi_{n_\pm}\rangle$ turns to
\begin{eqnarray}
	J_+|\phi_{n_+,n_-}\rangle&=&\sqrt{(n_++1)n_-}|\phi_{n_++1,n_--1}\rangle,\\
	J_-|\phi_{n_+,n_-}\rangle&=&\sqrt{n_+(n_-+1)}|\phi_{n_+-1,n_-+1}\rangle,\\
	J_3|\phi_{n_+,n_-}\rangle&=&\frac{1}{2}(n_+-n_-)|\phi_{n_+,n_-}\rangle.
\end{eqnarray}
From the  helicity occupation numbers $n_\pm$ representation, we introduce new set of  representation   label 
by the paire of  values $(j,m)$, which we define here as follows
\begin{eqnarray}
	n_++n_-=n=2j,\,\,\,\,\,\,\,\,\,\,\, n_+-n_-=2m,
\end{eqnarray}
such as
\begin{eqnarray}
	n_+=j+m,\,\,\,\,\,\,\,\,n_-=j-m,
\end{eqnarray}
where $j=0,\frac{1}{2},1,\frac{3}{2}...$ for  $n=0,1,2,3...$ and $-j\leq m\leq j$.
This representation is characterised by the single integer or half-integer number $j$, known as the
spin of that representation, whereas the states within that representation of given spin $j$ are distinguished
by their $J_3$ eigenvalue $m$ lying between $(-j)$ and $j$ in integer steps, with the above matrix elements for
the action of the two other operators  $J_\pm$ of the $SU(2)$ algebra.

Therefore we  define  a certain $SU(2)$ representation
as a finite $n + 1 = 2j + 1$ dimensional subspace of the complete Hilbert  space defined as follows
\begin{eqnarray}
	\mathcal{H}_j=\{|\phi_{n_\pm}\rangle=|\phi_{j,m}\rangle: 2j\in \mathbb{N},\,\,\,\,m=-j...j\}.
\end{eqnarray}
These states satisfy the properties of orthogonality and completeness 
\begin{eqnarray}
	\langle \phi_{j,m}|\phi_{j,m}\rangle=\delta_{mn},\,\,\,\, \sum_{m=0}^\infty|\phi_{j,m} \rangle\langle \phi_{j,m}|=\mathbb{I}.
\end{eqnarray}

The above action of the operators $J_\pm$ and $J_3$  then read  as
\begin{eqnarray}
	J_+|\phi_{j,m}\rangle&=&\sqrt{(j-m)(j+m+1)}|\phi_{j,m+1}\rangle,\\
	J_-|\phi_{j,m}\rangle&=& \sqrt{(j+m)(j-m+1)} |\phi_{j,m-1}\rangle,\\
	J_3|\phi_{j,m}\rangle&=&  m |\phi_{j,m}\rangle.
\end{eqnarray}
In conclusion, we have in this manner
recovered all finite dimensional representations of the symmetry algebra SU(2), which is also the algebra
of the symmetry group SO(3) of rotations in three dimensional Euclidean space. The above formula for the
action of $J_\pm$ and $J_3$ in a given spin $j$ representation are valid as such in full generality, independently of the
system in which these symmetries may be realised. Any rotationally invariant system in three dimensions
will find its quantum states organised according to these spin representations of $SU(2)$. But it is matter of
experiment to determine which spin values are realised for a specific physical system. For example, that
the electron has spin $\frac{1}{2}$ may only be determined experimentally.
As an illustration, consider the value $n= 1$ or $j = \frac{1}{2}$, namely the first excited state of the present
system. It is thus doubly degenerated, with
\begin{eqnarray}
	J_3|\phi_{\frac{1}{2}\,\frac{1}{2}}\rangle&=&\frac{1}{2}|\phi_{\frac{1}{2},\frac{1}{2}}\rangle,\,\,\, 
	J_+|\phi_{\frac{1}{2},\frac{1}{2}}\rangle=0,\,\,\, J_-|\phi_{\frac{1}{2},\frac{1}{2}}\rangle=\frac{1}{2}|\phi_{\frac{1}{2},-\frac{1}{2}}\rangle,\\
	J_3|\phi_{\frac{1}{2},-\frac{1}{2}}\rangle&=&-\frac{1}{2}|\phi_{\frac{1}{2},\frac{1}{2}}\rangle,\,\,\,
	J_+|\phi_{\frac{1}{2},-\frac{1}{2}}\rangle=\frac{1}{2}|\phi_{\frac{1}{2},\frac{1}{2}}\rangle,\,\,\,\,
	J_-|\phi_{\frac{1}{2},-\frac{1}{2}}\rangle=0,
\end{eqnarray}
and the associations  $|\phi_{j=\frac{1}{2},m=\frac{1}{2}}\rangle$= $|\phi_{n_+=1, n_-=0}\rangle$,
$|\phi_{j=\frac{1}{2},m =-\frac{1}{2}}\rangle= |\phi_{n_+=0, n_-=1}\rangle $. In other
words, in this two dimensional bases, the matrix representation of the operators $J_3$ and $J_\pm$ is given precisely
by the Pauli matrices. Clearly, the ground state of the system, $|\phi_{j=0, m=0}\rangle=|\phi_{n_+=0, n_-=0}\rangle$,
corresponds to a trivial representation of the SU(2) algebra for which $J_3=0$ and $J_\pm=0$.

Based on  what followed, the action of operators  $I(t), L_z$ are given by
\begin{eqnarray}
	I(t)|\phi_{j,m}\rangle&=&2\kappa(j+1) |\phi_{j,m}\rangle:,\\
	L_z|\phi_{j,m}\rangle&=& \frac{m}{2}|\phi_{j,m}\rangle,
\end{eqnarray}
and the expectation values of the Hamiltonian is given by
\begin{eqnarray}
	\langle \phi_{jm}| H(t)|\phi_{jm}\rangle&=&\frac{(j+1)}{\kappa}\left(M\dot{\rho}^2+\frac{\kappa^2}{M\rho^2}+M\Omega^2\rho^2\right)
	-m\omega_c-\frac{q^2E^2}{2M\omega}.
\end{eqnarray}

By successively applying $J_+$ on the ground state $|\phi_{j,-j}\rangle$, we generate the state $|\phi_{j,m}\rangle$ of the system as follows
\begin{eqnarray}
	|\phi_{j,m}\rangle=\sqrt{\frac{(j-m)!}{(j+m)!(2j)!}}(J_+)^{j+m}|\phi_{j,-j}\rangle,
\end{eqnarray}
where
\begin{eqnarray}
	J_-|\phi_{j,-j}\rangle=0.
\end{eqnarray}

\subsection{SU(2) coherent states}
The $SU(2$) coherent states for this model  denoted  by $|\phi_{j,\zeta}\rangle$
are obtained from  action of rotational group Lie on the Lowest weight state $|\psi_{j,-j}\rangle$
\begin{eqnarray}
	|\phi_{j,\zeta}\rangle&=& e^{\zeta J_++\zeta^* J_-}|\phi_{j,-j}\rangle\cr
	&=&\left(1+|\zeta|^2\right)^{-j}\sum_{m=0}^{2j}\left[\frac{(2j)!}{(m)!(2j-m)!}\right]^{\frac{1}{2}}\zeta^{m}|\phi_{j,m}\rangle,
\end{eqnarray}
where $\zeta=e^{i\phi}\tan\frac{\theta}{2}$ defined the complex plane which 
can been seen as a stereographic projection of sphere $(S^2)$ onto $\mathbb{C}$ of unit
vectors $(\theta,\phi)$.\\
These are normalized but are
non-orthogonal states and sastify the completeness relations
\begin{eqnarray}
	\langle \phi_{j,\zeta_1}|\phi_{j,\zeta_2}\rangle&=& (1+|\zeta_1|)^{-j}(1+|\zeta_2|)^{-j}(1+\zeta_1^*\zeta_2)^{2j},\\
	\int d\mu_j(\phi_{j,\zeta}\rangle\langle \phi_{j,\zeta}|&=&\sum_{m=0}^{2j}|\phi_{j,m}\rangle\langle \phi_{j,m}|=\mathbf{I},
\end{eqnarray}
$d\mu_j(j,\zeta)=\frac{2j+1}{\pi}\frac{d^2\zeta}{(1+|\zeta|^2)^2}$.\\
For  an   arbitrary states $|\Phi\rangle=\sum_{m=0}^{2j}c_m|\psi_{j,m}\rangle$    of  $\mathcal{H}_j$, one can use the above 
states to construct  in this space an analytical function $f(\zeta)$ such as
\begin{eqnarray}
	f(\zeta)&=&(1+|\zeta|^2)^j\langle \phi_{j,\zeta}^*|\Phi\rangle,\\
	&=&c_m\sum_{m=0}^{2j}\left[\frac{(2j)!}{(m)!(2j-m)!}\right]^{\frac{1}{2}}\zeta^{j+m}
\end{eqnarray}
The expansion of $|\Phi\rangle$ on the $SU(2)$ coherent states bases is
\begin{eqnarray}
	|\Phi\rangle&=&\int d\mu_j(j,\zeta)(1+|\zeta|^2)^{-j}f(\zeta^*)|\phi_{j,\zeta}\rangle,\\
	\langle \Phi|\Phi\rangle&=& \int d\mu_j(j,\zeta)(1+|\zeta|^2)^{-j}|f(\zeta^*)|^2<\infty.
\end{eqnarray}
\subsection{SU(2) photon added coherent states}
The Photon-added $SU(2)$ coherent states are obtained by successive application of the raising operator $J_+$ on $SU(2)$  coherent states
\begin{eqnarray}
	|\phi_{j\zeta,p}\rangle= \mathcal{W}_p^j(|\zeta|) (J_+)^p |\phi_{j\zeta}\rangle
\end{eqnarray}
Making use of the expression
\begin{eqnarray}
	(J_+)^p|\phi_{j,m}\rangle=\left[\frac{(m+p)!(2j-m)!}{(m)!(2j-m-p)!}\right]^{\frac{1}{2}}|\phi_{j,m+p}\rangle,
\end{eqnarray}
we finally obtain
\begin{eqnarray}
	|\phi_{j\zeta,p}\rangle&=& \mathcal{W}_p^j(|\zeta|)\sum_{m=0}^{2j-p}\left[\frac{(2j)!(m+p)!}{m!(2j-m-p)!}\right]^{\frac{1}{2}}
	\zeta^{m}|\phi_{j,m+p}\rangle
\end{eqnarray}
%
Using the above  expression, we obtain the normalization 
\begin{eqnarray}
	\langle \phi_{j\zeta,p}   |\phi_{j\zeta,p}\rangle=1,
\end{eqnarray}
from which results the normalization constant 
\begin{equation}
	\mathcal{W}_p^j(|\zeta|)=\left(\frac{\Gamma(1+2j)\Gamma(1+p){}_2\!F_1 (1+p,-2j+p;1;-|z|^2)}{\Gamma(1+2j-p)}\right)^{-1/2}
\end{equation}
The completeness relation is given as 
\begin{eqnarray}\label{menx4}
	\int d\mu(\zeta,\zeta^*)|\phi_{j,\zeta}\rangle \langle\phi_{j,\zeta}|=\mathbb{I}_j^p=
	\sum_{m=0}^{2j} |\phi_{j,m+p}\rangle\langle\phi_{j,m+p}|.
\end{eqnarray}
where  the integration measure  $d\mu(\zeta,\zeta^*)$  is given by
\begin{eqnarray}
	d\mu(\zeta,\zeta^*)=V_j^p(|\zeta|) \frac{d^2\zeta}{\pi}.
\end{eqnarray}
In case of $p=0$  the weight function  is reduced to $V_j^0(|\zeta|)=\frac{2j+1}{\pi(1+|\zeta|^2)^2}$.
Thus, in  order to resolve the identity operator in (\ref{menx4}), one should find  the general form of the  weight function
$V_j^p(|\zeta|)$. By means of a change of the complex variables in terms of polar coordinates $(\eta=re^{i\rho})$, using
the completeness of the states $|\phi_{j,m+p}\rangle$ and integrating on the angular variable $\rho$, we can use the resolution
of the identity operator (\ref{menx4}) to solve the following integral equation:
\begin{eqnarray}
	\int_0^\infty x^m \tilde V_j^p(x) dx&=&\frac{(m!)^2(2j-m-p)!}{(2j)!(m+p)!}
\end{eqnarray}
where $x=r^2$ and $\tilde V_j^p(x)=[\mathcal{W}_p^j(x)]^2V_j^p(x)$.
This is a Stieltjes moment problem and the function $V_j^p(x)$ may be found using the substitution
$m=s-1$
\begin{eqnarray}\label{menx5}
	\int_0^\infty x^{s-1}\tilde V_j^p(x) dx &=&\frac{[\Gamma(s)]^2\Gamma(2j+2-p-s)}{\Gamma(2j+1)\Gamma(m+p)}
\end{eqnarray}
It is useful to express the unknown function through the Meijer’s G-functions  and the Mellin inversion theorem.
Therefore using the definition of Meijer’s G-function, it  follows that
\begin{eqnarray}\label{menx6}
	\int_0^\infty dx x^{s-1}G_{p,q}^{n,m}
	\left(\alpha x\big|_{b_1,\cdots,\quad b_n,\quad b_{n+1},\quad \cdots, b_q}^{a_1,\cdots,\quad a_n,\quad a_{n+1},\quad \cdots, a_p}\right)\cr=
	\frac{1}{\alpha^s}
	\frac{\prod_{j=1}^n\Gamma(b_j+s)\prod_{j=1}^m\Gamma(1-a_j-s)}{\prod_{j=n+1}^q\Gamma(1-b_j-s)\prod_{j=m+1}^p\Gamma(a_j+s)}.
\end{eqnarray}
Comparing equations (\ref{menx5}) and (\ref{menx6}), we obtain that
\begin{eqnarray}
	\tilde V_j^p(x)= \frac{1}{\Gamma(2j+1)} G_{2,2}^{2,1}\left(x\big|_{0,\quad\,\,\, \,0\quad }^{-2j-1+p,\quad\quad p}\right).
\end{eqnarray}
The weight function $ V_j^p(|\zeta|)$ becomes
\begin{eqnarray}
	V_j^p(|\zeta|)=\frac{1}{[\mathcal{W}_p^j|\zeta|]^2\Gamma(2j+1)} G_{2,2}^{2,1}\left(x\big|_{0,\quad\,\,\, \,0\quad }^{-2j-1+p,\quad\quad p}\right)
\end{eqnarray}

For  an   arbitrary states $|\Phi\rangle=\sum_{m=0}^{2j}c_m|\phi_{k,m+n}\rangle$    in this space one may use the above 
states to construct  in this space an analytical function $f(\zeta)$ such as
\begin{eqnarray}
	f(\zeta)=\left(\mathcal{W}_p^{j}\right)^{-1}(|\zeta|)\langle \phi_{\zeta}^j|\Phi\rangle= \sum_{m=0}^{2j}c_m\left[\frac{(2j)!(m+p)!}{m!(2j-m-p)!}\right]^{\frac{1}{2}}\zeta^m
\end{eqnarray}

The photon count probability 
is given by
\begin{eqnarray}
	P_{j \zeta}(p)&=&|\langle \phi_{m+p}|\phi_{j\zeta,p}\rangle|^2,\cr
	&=&(\mathcal{W}_p^j)^2(|\zeta|)\sum_{m=0}^{2j-p}|\zeta|^{2m}\frac{(2j)!(m+p)!}{m!(2j-m-p)!}.                                            
\end{eqnarray}
\section{SU(1,1) symmetry and its coherent states for the model}

As we  pointed out in the previous section,  the eigenvalues of  the invariant operator $I(t)$ which  describes the stationary dynamic of the system
is degenerated. This degeneracy is related to the presence of the $SU(2)$ sysmmetry in the system. Whereas, the eigenvalues of the 
Hamiltonian  $H(t)$ of the system is not degenerated. This nondegeneracy can be described by the presence of another symmetry which is the $SU(1,1)$ symmetry.

Therefore, the $SU(1,1)$ algebra realization of the system  may be spanned by the following generators
\begin{eqnarray}
	K_0=\frac{1}{2}\left(a_+^\dag a_++a_-^\dag a_-+\mathbb{I}\right),\,\,\, K_+=a_+^\dag a_-^\dag,\,\,\,K_-=a_+a_-,
\end{eqnarray}
which satisfy 
\begin{eqnarray}
	[K_0,\pm K_\pm]=\pm K_\pm,\,\,\,\,\,\,\,[K_-,K_+]=2K_0.
\end{eqnarray}
.
The action of these operators on the Fock space states $\{|\psi_{k,n}\rangle\}$ is
\begin{eqnarray}
	K_+|\psi_{n_+,n_-}\rangle&=&\sqrt{(n_++1)(n_-+1)}|\psi_{n_++1,n_-+1}\rangle,\\
	K_-|\psi_{n_+,n_-}\rangle&=&\sqrt{n_+n_-}|\psi_{n_+-1,n_--1}\rangle,\\
	K_0|\psi_{n_+,n_-}\rangle&=&\frac{1}{2}(n_++n_-+1)|\psi_{n_++1,n_-+1}\rangle,\\
	K_+K_-|\psi_{n_+,n_-}\rangle&=&n_+n_-|\psi_{n_+,n_-}\rangle
\end{eqnarray}
From the  helicity occupation numbers $n_\pm$ representation, we introduce new set of  representation   label 
by the paire of  values $(k,m)$, which we define here as follows
\begin{eqnarray}
	k=\frac{1}{2}(n_+-n_++1),\,\,\,\, m=n_+,
\end{eqnarray}
where the parameters $k$ is  the Bargmann index which determine the representation of  $m$.

The action of operators $K_\pm$ and $K_0$ on the  space  span by $\{|k,m\rangle, m = 0, 1, 2, ...\}$ is
\begin{eqnarray}
	K_+|\psi_{k,m}\rangle&=&\sqrt{(m+1)(2k+m)}|\psi_{k,m+1}\rangle,\\
	K_-|\psi_{k,m}\rangle&=&\sqrt{m(2k+m-1)}|\psi_{k,m-1}\rangle,\\
	K_0|\psi_{k,m}\rangle&=&(k+m)|\psi_{k,m}\rangle.
\end{eqnarray}
We recover  from these actions of operators the standard $SU(1,1)$ irreducible representation 
\begin{eqnarray}\label{ec1}
	|k,m\rangle=\sqrt{\frac{\Gamma(2k)}{m!\Gamma(2k+m)}}(K_+)^m|k,0\rangle,
\end{eqnarray}
with 
\begin{eqnarray}
	K_-|k,0\rangle&=&0.
\end{eqnarray}
The corresponding Barut-Giraldello and the Perelomov coherent states  are given by
\begin{eqnarray}
	|\psi_{k,z}\rangle&=&\sqrt{\frac{|z|^{2k-1}}{I_{2k-1}(2|z|)}}\sum_{m=0}^\infty\frac{z^m}{\sqrt{m!\Gamma(m+2k)}}|\psi_{k,m}\rangle,\\
	|\psi_{k,\eta}\rangle&=&(1-|\eta|^2)^k\sum_{m=0}^\infty\sqrt{\frac{\Gamma(2k+m)}{m!\Gamma(2k)}}\eta^m|\psi_{k,m}\rangle.
\end{eqnarray}
From these coherent states, we aim to construct the corresponding the photon added coherent and  the nonlinear coherent states 
\subsection{Photon added coherent states}
\subsubsection{Perelomov photon added  coherent states}
The photon  added coherent states associated to Perelomov  $SU(1,1)$  coherent states are  obtained by successive action of the raising operator 
$K_+$ on  the Perelomov  $SU(1,1)$  coherent states
\begin{eqnarray}
	|\psi_{k\eta,l}\rangle&=&\mathcal{N}_l(K_+)^l|\psi_{k\eta}\rangle,\cr
	&=& \mathcal{N}_l(1-|\eta|^2)^k\sum_{m=0}^\infty\sqrt{\frac{\Gamma(2k+m)}{m!\Gamma(2k)}}\eta^m
	(K_+)^n|\psi_{k,m}\rangle.                       
\end{eqnarray}
Making use of the expression
\begin{eqnarray}
	(K_+)^l|\psi_{k\eta}\rangle&=&\sqrt{\frac{(m+l)!(m+2k+l-1)!}{m!(2k+m-1)!}}|\psi_{k,m+l}\rangle\cr
	&=& \sqrt{\frac{\Gamma(m+l+1)\Gamma(m+2k+l)}{m!\Gamma(2k+m)}}|\psi_{k,m+l}\rangle,
\end{eqnarray}
we then obtain
\begin{eqnarray}
	|\psi_{k\eta,l}\rangle&=&\mathcal{N}_l(|\eta|)\sum_{m=0}^\infty\frac{\eta^m}{\sqrt{F_l(k,m)}}                       
	|\psi_{k,m+l}\rangle,
\end{eqnarray}
where 
\begin{eqnarray}
	F_l(k,m)&=& \left[\frac{\Gamma(m+l+1)\Gamma(m+2k+l)}{(m!)^2\Gamma(2k)}\right]^{-1}\\
	\mathcal{N}_l(|\eta|)&=&\frac{(1-|\eta|^2)^k}{\sqrt{\langle\psi_{k\eta}|(K_-)^l(K_+)^l|\psi_{k,\eta}\rangle}},
\end{eqnarray}
The non-orthogonality is expressed as
\begin{eqnarray}
	\langle \psi_{k\eta',l'}|\psi_{k\eta,l}\rangle&=&\mathcal{N}_l(|\eta|)\mathcal{N}_{l'}(|\eta'|)
	\sum_{m=0}^\infty\frac{\eta^m\eta*^{m'}}{\sqrt{F_l(k,m)F_{l'}(k,m')}}                 
	\langle\psi_{k,m'+l'}|\psi_{k,m+l}\rangle,\cr
	&=&\mathcal{N}_l(|\eta|)\mathcal{N}_{l'}(|\eta'|) {\eta'}*^{l-l'}\cr&&\times
	\sum_{m=0}^\infty \frac{(\Gamma(m+2l-l'+1)\Gamma(m+2l-l'+2k))^{1/2}}{\sqrt{m!\Gamma(2k)}}\cr&&\times
	\frac{(\Gamma(m+l+1)\Gamma(m+2k+l))^{1/2}}{\sqrt{m!\Gamma(2k)}}
	\frac{ ({\eta'}*\eta)^m}{m!}
\end{eqnarray}
The overcompletion property of these states is
\begin{eqnarray}\label{menx1}
	\int_{\mathbb{C}} d\nu(\eta,\eta^*)|\psi_{k\eta,l}\rangle \langle\psi_{k\eta,l}|=\mathbb{I}_k^l=
	\sum_{m=0}^\infty |\psi_{k,m+l}\rangle\langle\psi_{k,m+l}|,
\end{eqnarray}
where the determination of the integration measure   $d\nu(\eta,\eta^*)$  guarantees the resolution of the unity.
To this end, we assume the existence of a positive weight function $W_k^l(|\eta|)$ such that
\begin{eqnarray}
	d\nu(\eta,\eta^*)=\frac{d^2\eta}{\pi}W_k^l(|\eta|),
\end{eqnarray}
where $\frac{d^2\eta}{\pi}$ the usual Lebesgue measure. For $l=0$ this weight function is reduced to the ordinary Perelomov one,
\begin{eqnarray}
	W_k^0(|\eta|)=\frac{2k-1}{\pi(1-|\eta|^2)^2}.
\end{eqnarray}
By means of a change of the complex variables in terms of polar coordinates $\eta=re^{i\varrho}$ where 
$r\in \mathbb{R}_+$,\,\,\,$\varrho \in [0,2\pi)$,\,\,\, and 
$d^2(\eta)=rdrd\varrho$,  the equation (\ref{menx1}) becomes
\begin{eqnarray}\label{menx2}
	\sum_{m'm=0}\left[\frac{1}{\sqrt{F_l(k,m)F_{l}(k,m')}}\int_0^\infty dr r^{1+m'+m}W_k^l(r^2)\mathcal{N}_l^2(r^2)
	\int_0^{2\pi}\frac{d\varrho}{\pi}e^{i(m-m')\varrho}\right]\cr\times|\psi_{k,m+l}\rangle\langle\psi_{k,m'+l}|=\mathbb{I}_k^l
\end{eqnarray}
By performing the angular integration, i.e
\begin{equation}
	\int_0^{2\pi}\frac{d\varrho}{\pi}e^{i(m-m')\varrho}=2\delta_{mm'},
\end{equation}
the resolution of the identity operator (\ref{menx2}) is
\begin{eqnarray}\label{menx2}
	2\sum_{m=0}\left[\frac{1}{F_l(k,m)}\int_0^\infty dr r^{1+2m}W_k^l(r^2)\mathcal{N}_l^2(r^2)\right]
	|\psi_{k,m+l}\rangle\langle\psi_{k,m'+l}|=\mathbb{I}_k^l.
\end{eqnarray}
Setting the weight function such as
\begin{eqnarray}\label{menx3}
	W_k^l(r^2)=\frac{1}{\mathcal{N}_l^2(r^2)}g_k^l(r^2),
\end{eqnarray}
and using the completeness of the states $ |\psi_{k,m+l}\rangle$ and taking $x=r^2$, we have the following integral to solve
\begin{eqnarray}
	\int_0^\infty dx x^m g_k^l(x)= F_l(k,m)=\frac{(m!)^2\Gamma(2k)}{\Gamma(m+l+1)\Gamma(m+2k+l)}.
\end{eqnarray}
To solve  this integral, one should find the function $g_k^l(x)$. To do so, we  proceed as folows.
Instead of  solving this integral for $g_k^l(x)$, we shall study its existence of solutions.
The equation (\ref{menx3}) is the well-known Stieltjes moment problem, and this integral equation does not admit a
general solution. In fact, whether a solution exists or not and the form of this solution in the positive
case depends on the form of the box function. For solving this integral equation,
we use the Fourier transforms method by multiplying Eq. (\ref{menx3}) by $\frac{(iy)^m}{m!}$ and summing over $m$ yields

\begin{eqnarray}
	\int_0^\infty dx e^{iyx} g_k^l(x)= \sum_{m=0}^\infty F_l(k,m) \frac{(iy)^m}{m!}=\large{\bar g}_k^l(y) .
\end{eqnarray}
In the case where the series defining the function $\large{\bar g}_k^l(y)$ above converges, the inverse Fourier transforms
reads
\begin{eqnarray}
	g_k^l(x)=\frac{1}{2\pi}\int_{-\infty}^{+\infty}e^{-ixy}\large{\bar g}_k^l(y)dy.
\end{eqnarray}
Finally, the weight function $W_k^l(x)$ allowing for a resolution of the identity operator,  is written as
\begin{eqnarray}
	W_k^l(x)=\frac{1}{2\pi \mathcal{N}_l^2(x)}\int_{-\infty}^{+\infty}e^{-ixy}\large{\bar g}_k^l(y)dy.
\end{eqnarray}

For  an   arbitrary states $|\Phi\rangle=\sum_{m=0}^{\infty}c_m|\psi_{k,m+n}\rangle$    in this space one can use the above 
states to construct  in this space an analytical function $f(\eta)$ such as
\begin{eqnarray}
	f(\eta)&=&\mathcal{N}_l^{-1}(|\eta|)\langle \psi_{k,\eta}|\Phi\rangle,\\
	&=&c_m\sum_{m=0}^\infty\frac{(\eta*)^m}{\sqrt{F_n(k,m)}}.
\end{eqnarray}

The photon distribution 
is given by
\begin{eqnarray}
	P_{k\eta}(n)&=&|\langle \psi_{l+m}|\psi_{k\eta,l}\rangle|^2,\cr
	&=&\mathcal{N}_l^2(|\eta|)\sum_{m=0}^\infty\frac{|\eta|^{2m}}{F_l(k,m)}.                                              
\end{eqnarray}

\subsubsection{Barut-Giraldello photon added  coherent states}

Photon added of Barut-Giraldello coherent states can be obtained by repeated application of the raising operator $K_+$ on the Barut-Giraldello
coherent states
\begin{eqnarray}\label{ec3}
	|\psi_{kz,n}\rangle&=&\mathcal{M}_n(K_+)^n|\psi_{kz}\rangle,\cr
	&=& \mathcal{M}_n\sqrt{\frac{|z|^{2k-1}}{I_{2k-1}(2|z|)}}\sum_{m=0}^\infty\frac{z^m}{\sqrt{m!\Gamma(m+2k)}}
	(K_+)^n|\psi_{k,m}\rangle,\cr
	&=&\mathcal{M}_n(|z|)\sum_{m=0}^\infty\frac{z^m}{\sqrt{\rho_n(k,m)}}                       
	|\psi_{k,m+n}\rangle,
\end{eqnarray}
where we have used the notation
\begin{eqnarray}
	\rho_n(k,m)&=&\frac{[\Gamma(m+1)]^2[\Gamma(m+2k)]^2}{\Gamma(m+n+1)\Gamma(m+n+2k)},\\
	\mathcal{M}_n(|z|)&=&\frac{\sqrt{\frac{|z|^{2k-1}}{I_{2k-1}(2|z|)}}}{\sqrt{\langle\psi_{kz}|(K_-)^n(K_+)^n|\psi_{kz}\rangle}}.
\end{eqnarray}
The non-orthogonality condition is realised as follows
\begin{eqnarray}
	\langle \psi_{k,z_2,n'}|\psi_{k,z_1,n}\rangle= \mathcal{M}_n'(|z_2|)\mathcal{M}_n(|z_1|) 
	\frac{z_1^m(z_2^*)^{m'}}{\sqrt{\rho_n(k,m)\rho_{n'}(k,m')}}                 
	\langle \psi_{k,m'+n'}|\psi_{k,m+n}\rangle.
\end{eqnarray}
Due to the orthogonality relation of the number vectors (\ref{ec1}), it follows that
\begin{eqnarray}
	\langle \psi_{k,z_2,n'}|\psi_{k,z_1,n}\rangle&=&  \mathcal{M}_n'(|z_2|)\mathcal{M}_n(|z_1|)
	(z_2^*)^{n-n'}\frac{[\Gamma(n+1)]^2[\Gamma(n+2k)]^2}{\Gamma(n-n'+1)\Gamma(n-n'+2k)\Gamma (2k)}
	\times \cr&& {}_2\!F_3 (n+1,n+2k;n-n'+1,n-n'+2k,2k;z_2^*z_1),
\end{eqnarray}
where $n$ and $n'$ are positive integers, $n\geq n'$ and $ {}_2\!F_3 (\cdots;z_2^*z_1)$
is the generalized hypergeometric series (see appendix )

The resolution of unity operator in this space is
\begin{eqnarray}\label{ec12}
	\int_{\mathbb{C}}\frac{d^2 z}{\pi}Y_k^n(|z|)|\psi_{kz,n}\rangle \langle\psi_{kz,n}|=
	\mathbb{I}_k^n= \sum_{m=0}^\infty |\psi_{k,m+n}\rangle\langle\psi_{k,m+n}|.
\end{eqnarray}
The overcompleteness is satisfy
if one can find a function  $ Y_k^n(|z|)$. At the limit $n= 0$, this weight function must lead to the weight function of the ordinary
Barut-Giraldello coherent states,
\begin{eqnarray}
	Y_k^0(|z|)=2K_{2k-1}(2|z|)I_{2k-1}(2|z|).
\end{eqnarray}
For $n\neq 0$, by substituting equation (\ref{ec3}) into equation (\ref{ec12}) we obtain
\begin{eqnarray}\label{ec4}
	\int_{\mathbb{C}}\frac{d^2 z}{\pi}Y_k^n(|z|) [\mathcal{M}_n(|z|)]^2 \sum_{m,m'=0}^\infty 
	\frac{(z^*)^{m'}z^m}{\sqrt{\rho_{n}(k,m')\rho_{n}(k,m)}}|\psi_{k,m+n}\rangle\langle\psi_{k,m'+n}|=\mathbb{I}_k^n.
\end{eqnarray}
Then, it is obvious that the weight function $Y_k^n (|z|)$ must have the following structure
\begin{equation}
	Y_k^n (|z|)=\frac{1}{[\mathcal{M}_n(|z|)]^2 }|z|^{2m} g_k^m (|z|)
\end{equation}
By considering the following transformation $z=re^{i\theta}$,\,\,\, $r\in \mathbb{R}_+$,\,\,\,$\theta\in [0,2\pi)$,\,\,\,
$d^2z=rdrd\theta$ and after performing the angular integration, i.e
\begin{equation}
	\int_0^{2\pi}\frac{d\theta}{\pi}e^{i(m-m')\theta}=2\delta_{mm'},
\end{equation}
equation (\ref{ec4}) becomes
\begin{eqnarray}
	2\sum_{m=0}^\infty\left[\frac{1}{\rho_{n}(k,m)}\int_0^\infty dr r^{2m+2n+1}g_k^n(r^2)\right]
	|\psi_{k,m+n}\rangle\langle\psi_{k,m+n}|=\mathbb{I}_k^n.
\end{eqnarray}
When we perform the variable change $r^2=x$ and  $n + m=s-1$, the integral from the above equation is called the Mellin
transform (see Appendix)
\begin{equation}\label{ec5}
	\int_0^\infty dx x^{s-1}g_k^n(x)=\rho_n(k,s-n-1)=\frac{[\Gamma(s-n)]^2[\Gamma(s-n+2k-1)]^2}{\Gamma(s)\Gamma(s+2k-1)}
\end{equation}
Using the definition of Meijer’s G-function, it  follows that
\begin{eqnarray}\label{ec6}
	\int_0^\infty dx x^{s-1}G_{p,q}^{m,n}
	\left(\alpha x\big|_{b_1,\cdots,\quad b_m,\quad b_{m+1},\quad \cdots, b_q}^{a_1,\cdots,\quad a_n,\quad a_{n+1},\quad \cdots, a_p}\right)\cr=
	\frac{1}{\alpha^s}
	\frac{\prod_{j=1}^m\Gamma(b_j+s)\prod_{j=1}^n\Gamma(1-a_j-s)}{\prod_{j=m+1}^q\Gamma(1-b_j-s)\prod_{j=n+1}^p\Gamma(a_j+s)}.
\end{eqnarray}
Comparing equations (\ref{ec5}) and (\ref{ec6}), we obtain that
\begin{eqnarray}
	g_k^n(x)= G_{2,4}^{4,0}\left(x\big|_{-n,\,\,\,\,-n,\quad 2k-1-n,\quad  2k-1-n}^{0,\,\,\,\,\,\;\;\,2k-1}\right).
\end{eqnarray}
Then, the weight function becomes
\begin{eqnarray}
	Y_k^n(|z|)=\frac{1}{[\mathcal{M}_n(|z|)]^2 }
	G_{2,4}^{4,0}\left(|z|^2\big|_{0,\quad\,\,\, \,0,\quad\quad\,\,\,\,\,\, 2k-1,\quad\quad  2k-1}^{n,\,\,\,\,\,2k-1+n}\right)
\end{eqnarray}
Finally, the resolution of unity can be written in an exhaustive manner
\begin{eqnarray}\label{ec2}
	\int_{\mathbb{C}}\frac{d^2 z}{\pi}\frac{1}{[\mathcal{M}_n(|z|)]^2 }
	G_{2,4}^{4,0}\left(|z|^2\big|_{0,\quad\,\,\, \,0,\quad\quad\,\,\,\,\,\, 2k-1,\quad\quad  2k-1}^{n,\,\,\,\,\,2k-1+n}\right)|\psi_{k,m+n}\rangle\langle\psi_{k,m+n}|
	=\mathbb{I}_k^n.
\end{eqnarray}

For  an   arbitrary states $|\Psi\rangle=\sum_{m=0}^{\infty}c_m|\psi_{k,m+n}\rangle$    in this space one can use the above 
states to construct  in this space an analytical function $f(z)$ such as
\begin{eqnarray}
	f(z)=\mathcal{M}_n^{-1}(|z|)\langle \psi_{z}^k|\Psi\rangle= \sum_{m=0}^\infty c_m\frac{z^m}{\sqrt{\rho_n(k,m)}}.
\end{eqnarray}
The photon distribution 
is given by
\begin{eqnarray}
	P_{k z}(n)&=&|\langle \psi_{n+m}|\psi_{kz,n}\rangle|^2,\cr
	&=&\mathcal{M}_n^2(|z|)\sum_{m=0}^\infty\frac{|z|^{2m}}{\rho_n(k,m)}.                                            
\end{eqnarray}

\section{ SU(1,1) coherent states from factorisation of $U(1)$ wavefunction }
To construct  the $SU(1,1)$ coherent states for this system whose
eigenfunction is  expressed in terms of the generalized Laguerre functions, we firstly  factorise this eigenfunction 
to discover the hidden $su(1,1)$  symmetry of the system. Secondly, we construct the corresponding coherent states and deduce the related properties. 
Finally,  from the $SU(1,1)$-coherent states constructed we constructed the associated nonlinear and photon added coherent states.

\subsection{The hidden dynamical $su(1,1)$ algebra}
We construct here   the raising and lowering operators from the eigenfunctions of this system which generate the hidden $su(1,1)$ algebra. 
Since  the eigenfunctions of the invariant operator  and  the Hamiltonian   are expressed in terms of the generalized Laguerre
functions $L_n^{\ell}(u)$ with $\ell>0$.
It is important to review some useful properties  related to this  special function that will be used to generate the symmetry operators.
Thus, the generalised Laguerre polynomials $L_n^{\ell}(u)$  are defined  as \cite{169}
\begin{eqnarray}
	L_n^{\ell}(u)&=&\frac{1}{n!}e^u u^{-\ell}\frac{d^n}{du^n}(e^{-u}u^{n+\ell}).
\end{eqnarray}
For $\ell=0,\,\,\,L_n^0(u)=L_n(u)$,  and \,\,\,for $n=0,\,\,\,L_0^{\ell}(u)=1$. The generating functions corresponding to the associated
Laguerre polynomials are
\begin{eqnarray}
	\frac{e^{\frac{uz}{z-1}}}{(1-z)^{1+\ell}}&=&\sum_{n=0}^\infty L_n^{\ell}(u)z^n,\,\,\,\,\,\,|z|<1,\label{lan32}\\
	J_{\ell}\left(2\sqrt{uz}\right)e^z(uz)^{-\frac{\ell}{2}}&=&\sum_{n=0}^\infty\frac{z^n}{\Gamma(n+\ell+1)}L_n^{\ell}(u),\label{lan33}
\end{eqnarray}
where the $ J_\kappa(x)$ is the ordinary Bessel function of $\kappa$-order.\\
The orthogonality relation is 
\begin{eqnarray}
	\int_0^{\infty}du e^{-u}u^{\ell}L_n^{\ell}(u)L_m^{\ell}(u)= \frac{\Gamma(\ell+n+1)}{n!} \delta_{nm}                                                     
\end{eqnarray}
The generalised Laguerre polynomials satisfy the following differential equation
\begin{equation}
	\left[u\frac{d^2}{du^2}+(\ell-u+1)\frac{d}{du}+n\right]L_n^{\ell}(u)=0,
\end{equation}
and the recurrence relations
\begin{eqnarray}
	(n+1)L_{n+1}^{\ell}(u)-\left(2n+\ell+1-u\right)L_n^{\ell}(u)+\left(n+\ell\right)L_{n-1}^{\ell}(u)&=&0,\label{lan27}\\
	u\frac{d}{du}L_n^{\ell}(u)-nL_n^{\ell}(u)+(n+\ell)L_{n-1}^{\ell}(u)&=&0.\label{lan28}
\end{eqnarray}
In review of these, we rewritte the  eigenfunction of the invariant operator (\ref{lan26}) in the form
\begin{equation}
	\phi_n^{\ell}(u)=N(\rho,\alpha)\sqrt{\frac{n!}{\Gamma(n+\ell+1)}}u^{\frac{\ell}{2}}e^{-\frac{\beta}{2}u}L_n^{\ell}(u),
\end{equation}
where $u=\frac{\kappa}{\rho^2}r^2$, $N(\rho,\theta)=(-)^n\sqrt{\frac{\kappa}{\pi\rho^2}}e^{i\ell\theta}$,\,\,\,\,
$\beta=1-iM\frac{\rho\dot{\rho}}{\kappa}$ and $\Gamma(n)=(n-1)!$.\\
Basing on the recurrence relation (\ref{lan27}) and (\ref{lan28}), we obtain the following equations
\begin{eqnarray}
	\left(-u\frac{d}{du}+\frac{\ell}{2}+n-\frac{\beta}{2}u\right)\phi_n^{\ell}(u)&=&\sqrt{n(n+\ell)}\phi_{n-1}^{\ell}(u),\\
	\left(u\frac{d}{du}+\frac{\ell}{2}+n-\frac{\tilde \beta}{2}u+1\right)\phi_n^{\ell}(u)&=&\sqrt{(n+1)(n+\ell+1)}\phi_{n+1}^{\ell}(u),
\end{eqnarray}
where $\tilde \beta=2-\beta$.
For the sake of simplicity we define the raising operator $K_+$ and the lowering operator $K_-$ for the generalised Laguerre functions as
\begin{eqnarray}
	K_-=\left(-u\frac{d}{du}+\frac{\ell}{2}+n-\frac{\beta}{2}u\right),\\
	K_+=\left(u\frac{d}{du}+\frac{\ell}{2}+n-\frac{\tilde \beta}{2}u+1\right),
\end{eqnarray}
and  hence obtain
\begin{eqnarray}
	K_-\phi_n^{\ell}(u)&=&\sqrt{n(n+\ell)}\phi_{n-1}^{\ell}(u),\\
	K_+\phi_n^{\ell}(u)&=&\sqrt{(n+1)(n+\ell+1)}\phi_{n+1}^{\ell}(u).
\end{eqnarray}
By multiplying  both side of the latter  equations  by the factor $e^{i\gamma_{n,\ell}(t)}$  we have
\begin{eqnarray}
	K_-\psi_n^{\ell}(u)&=& \sqrt{n(n+\ell)}\psi_{n-1}^{\ell}(u),\label{lan31}\\
	K_+\psi_n^{\ell}(u)&=&\sqrt{(n+1)(n+\ell+1)}\psi_{n+1}^{\ell}(u).
\end{eqnarray}
By successively applying $K_+$ on the ground state $\psi_0^{\ell} (u)$, we generate the eigenfunction  $\psi_n^{\ell} (u)$ of the system 
as follows
\begin{eqnarray}
	\psi_n^{\ell}(u)&=&\sqrt{\frac{\Gamma(1+\ell)}{n!\Gamma(n+\ell+1)}}(K_+)^n\psi_0^{\ell} (u),\\
\end{eqnarray}
where,
\begin{eqnarray}
	\psi_0^{\ell}(u)&=&\frac{N(\rho,\theta)}{\sqrt{\Gamma(\ell+1)}}u^{\frac{\ell}{2}}e^{-\frac{\beta}{2}u}e^{i\theta_{n,\ell}(t)},\\
	K_- \psi_0^{\ell}(u)&=&0.
\end{eqnarray}
One can also observe that the following relations are satisfied
\begin{eqnarray}
	K_+K_-\psi_n^{\ell} (u)&=&n(n+\ell)\psi_n^{\ell} (u),\\
	K_-K_+\psi_n^{\ell} (u)&=&(n+1)(n+\ell+1)\psi_n^{\ell} (u).
\end{eqnarray}
Now, to establish the dynamic group associated with the ladder operators $K_\pm$, we calculate the
commutator
\begin{equation}
	[K_-,K_+]\psi_n^{\ell}(u)=(2n+\ell+1)\psi_n^{\ell}(u).
\end{equation}
As a consequence, we can introduce the operator $K_0$ defined to satisfy
\begin{eqnarray}
	K_0\psi_n^{\ell}(u)=\frac{1}{2}(2n+\ell+1)\psi_n^{\ell}(u).
\end{eqnarray}
The operators $K_\pm$ and $K_0$  satisfy the following  commutation relations
\begin{eqnarray}
	[K_-,K_+]=2K_0,\,\,[K_0,K_\pm]=\pm K_\pm,
\end{eqnarray}
which can be recognized as commutation relation of the generators of a non-compact Lie
algebra $su(1,1)$. The corresponding Casimir operator for any irreducible representation is the identity times a number
\begin{eqnarray}
	K^2=K_0^2-\frac{1}{2}(K_+K_-+K_-K_+)=\frac{1}{4}(\ell+1)(\ell-1).
\end{eqnarray}
It satisfies
\begin{equation}
	[K^2,K_\pm]=[K^2,K_0]=0.
\end{equation}
Thus, a representation of $su(1, 1)$ algebra is determined by the single real positive number $\ell$, called the Bargmann index.\\
Now, with the properties of the generators $K_\pm$ and $K_0$ of the $su(1,1)$ algebra, we are in the position to construct the corresponding
coherent states  to this system.

We investigate in this section the $SU(1,1)$ coherent states by adopting  Barut-Girardello \cite{84} and Perelomov \cite{83} approaches. We examin for each
approach the resolution of unity and overlapping properties.
\subsection{Barut-Girardello coherent states}
\subsubsection{Construction}
Following the Barut and Girardello approach \cite{84},  $ SU(1,1)$ coherent states  are defined  to  be 
the eigenstates of the lowering generator $K_-$
\begin{equation}\label{lan29}
	K_-|\psi_{z}^\ell\rangle=z |\psi_{z}^\ell\rangle,
\end{equation}
where $z$ is an arbitrary complex number. Based on the completeness of the wavefunction such that $|\psi_n^{\ell}\rangle\langle \psi_n^{\ell}|=\mathbb{I}$, on
can represent the coherent states $|z,\ell\rangle$ as follows
\begin{eqnarray}\label{lan30}
	|\psi_{z,}^\ell\rangle&=& \sum_{n=0}^\infty \langle \psi_n^{\ell}|\psi_{z,}^\ell\rangle|\psi_n^{\ell}\rangle.
\end{eqnarray}
Acting the operator $K_-$ on the equation (\ref{lan30}) and then, using the equations (\ref{lan29}) and (\ref{lan31})
we have the following result
\begin{eqnarray}
	\langle \psi_n^{\ell}|\psi_{z,}^\ell\rangle=\frac{z}{\sqrt{(n+1)(n+\ell+1)}}\langle \psi_{n-1}^{\ell}|\psi_{z}^\ell\rangle.
\end{eqnarray}
After the recurrence procedure,  the formal equation becomes
\begin{equation}
	\langle \psi_n^{\ell}|\psi_{z,}^\ell\rangle=z^n\sqrt{\frac{\Gamma(1+\ell)}{n!\Gamma(n+\ell+1)}}\langle \psi_0^{\ell}|\psi_{z,}^\ell\rangle
\end{equation}
Referring to \cite{169}, the Gamma function is linked to the modified Bessel function $I_\mu(x)$ of order $\mu$ through the relation
\begin{equation}
	\sum_{n=0}^\infty\frac{x^{2n}}{n!\Gamma(n+\mu+1)}=\frac{I_\mu(2x)}{x^\mu}.
\end{equation}
Therefrom, by setting $x=z$  and  $\mu=\ell$,  we deduce the Barut-Girardello coherent states as fallows
\begin{eqnarray}
	|\psi_{z,}^\ell\rangle&=&\sqrt{\frac{|z|^{\ell}}{I_{\ell}(2|z|)}}\sum_{n=0}^\infty\frac{z^n}{\sqrt{n!\Gamma(n+\ell+1)}}|\psi_n^{\ell}\rangle,
	\label{lan34}\\
	\psi_z^{\ell}(u)&=&\frac{|z|^{\frac{\ell}{2}} N(\rho,\alpha)}{\sqrt{I_{\ell}(2|z|)}}
	\sum_{n=0}^\infty \frac{z^n}{\Gamma(n+\ell+1)}u^{\frac{\ell}{2}}e^{-\frac{\beta}{2}u}L_n^{\ell}(u)e^{i\gamma_{n,\ell}(t)}.
\end{eqnarray}
Taking the limit $t\rightarrow 0$ the phase factor $\gamma_{n,\ell} (t)\rightarrow 0$ and $ \psi_z^{\ell}(u)\rightarrow \phi_z^{\ell}(u)$.\\
However, in term of the generating function  (\ref{lan33}), the Barut-Girardello coherent  states can be written as follows
\begin{eqnarray}
	\psi_z^{\ell}(u)&=&\left(\frac{z}{|z|}\right)^{-\frac{\ell}{2}}\frac{ N(\rho,\theta) e^{z-\frac{\beta}{2}u} }{\sqrt{I_{\ell}(2|z|)}}
	J_{\ell}\left(2\sqrt{uz}\right)e^{i\gamma(t)}.
\end{eqnarray}
\subsubsection{Properties}
It is well-known  that the states (\ref{lan34}) are normalized but not orthogonal 
and satisfy the resolution of identity. Thus, we can see that the 
scalar product of two coherent states does not vanish
\begin{equation}
	\langle \psi_{z_1}^\ell|\psi_{z_2}^\ell\rangle=\frac{I_{\ell}(2\sqrt{ z_1^*z_2)}}{\sqrt{I_{\ell}(2|z_1|)|I_{\ell}(2|z_2|)}}.
\end{equation}
The overcompleteness relation reads  as follows
\begin{equation}
	\int d\mu(z,\ell)|\psi_{z}^\ell\rangle\langle \psi_{z}^\ell|=\sum_{n=0}^\infty|\psi_n^{\ell}\rangle\langle \psi_n^{\ell}|=\mathbb{I},
\end{equation}
with the measure
\begin{equation}
	d\mu(z,\ell)=\frac{2}{\pi}K_{\ell}(2|z|)I_{\ell}(2|z|)d^2z,
\end{equation}
where $d^2z=d(Re z)d(Im z)$ and $K_\nu(x)$ is the $\nu$-order modified Bessel function of the second kind.\\
For arbitrary  state $|\Phi\rangle=\sum_{n=0}^\infty c_n|\psi_n^{\ell}\rangle$ in the Hilbert space, one can construct  the analytic function $f(z)$
such that
\begin{eqnarray}
	f(z)=\sqrt{\frac{I_{\ell}(2|z|)}{|z|^{\ell}}}\langle \psi_{z}^\ell|\Phi\rangle=\sum_{m=0}^\infty\frac{c_m}{\sqrt{m!\Gamma(m+\ell+1)}}z^m.
\end{eqnarray}
On the Barut-Girardello coherent states (\ref{lan34}) one  can explicitly express the state $|\Phi\rangle$ as follows
\begin{equation}
	|\Phi\rangle=\int d\nu(z,\ell)\frac{({z^*})^{\frac{\ell}{2}}}{\sqrt{I_{\ell}(2|z|)}}f(z)|\psi_{z}^\ell\rangle,
\end{equation}
and  we have
\begin{equation}
	\langle \Phi|\Phi\rangle=\int d\mu(z,\ell)\frac{|z|^{\ell}}{I_{\ell}(2|z|)}|f(z)|^2<\infty.
\end{equation}
\subsection{Perelomov coherent states}
\subsubsection{Construction}
In analogy   to canonical   coherent states   construction, Perelomov $SU(1,1)$
coherent states $|\psi_{\eta}^\ell\rangle$ \cite{83} are obtained by acting  the displacement operator  $S(\xi)$ on the ground state 
$|\psi_0^{\ell}\rangle$
\begin{eqnarray}
	|\psi_{\eta}^\ell\rangle&=& S(\xi)|\psi_0^{\ell}\rangle,\cr
	&=& \exp\left(\xi K_+-\xi^*K_-\right)|\psi_0^{\ell}\rangle,
\end{eqnarray}
where $\xi\in\mathbb{C},$  such that $\xi= -\frac{\theta}{2}e^{-i\varphi}$, with $-\infty<\theta<+\infty$  and $0\leq\varphi\leq 2\pi$.\\
Using Baker-Campbell-Haussdorf relation we explicit the displacement as fallows
\begin{equation}\label{lan35}
	S(\xi)= \exp(\eta K_+)\exp(\zeta K_0)\exp(-\eta^* K_-),
\end{equation}
where $\eta=-\tanh(\frac{\theta}{2})e^{-i\varphi}$ and $\zeta=-2\ln\cosh|\xi|=\ln(1-|\eta|^2)$. By using this normal 
form of the  displacement operator  (\ref{lan35}), the standard Perelomov $SU(1,1)$ coherent states are  found to be
\begin{eqnarray}
	|\psi_{\eta}^\ell\rangle&=&(1-|\eta|^2)^{\ell+1}\sum_{n=0}^\infty\sqrt{\frac{\Gamma(n+\ell+1)}{n!\Gamma(\ell+1)}}\eta^n|\psi_n^{\ell}\rangle,\label{lan36}\\
	\psi_\eta^{\ell}(u)&=&N(\rho,\theta)\frac{(1-|\eta|^2)^{\ell+1}}{\sqrt{\Gamma(\ell+1)}}u^{\frac{\ell}{2}}e^{-\frac{\beta}{2}u}
	\sum_{n=0}^\infty \eta^n L_n^{\ell}(u) e^{i\gamma_{n,\ell}(t)}.
\end{eqnarray}
Taking the limit $t\rightarrow 0$ the phase factor $\gamma_{n,\ell} (t)\rightarrow 0$ and $ \psi_\eta^{\ell}(u)\rightarrow \phi_\eta^{\ell}(u)$.\\
In term of the generating function  (\ref{lan32}), the Perelomov coherent  states can be written as follows
\begin{equation}
	\psi_\eta^{\ell}(u)=N(\rho,\theta)\frac{(1-|\eta|^2)^{\ell+1}}{\sqrt{\Gamma(\ell+1)}}u^{\frac{\ell}{2}}e^{-\frac{\beta}{2}u}
	\frac{e^{\frac{u\eta}{\eta-1}}}{(1-\eta)^{1+\ell}}e^{i\gamma(t)}.
\end{equation}

\subsection{Properties}
The Perelomov $SU(1,1)$ coherent states as the Barut-Girardello coherent states are normalized states but not orthogonal 
\begin{equation}
	\langle \psi_{\eta_1}^\ell|\psi_{\eta_2}^\ell\rangle=\left[(1-|\eta_1|^2)(1-|\eta_2|^2)\right]^{\frac{\ell+1}{2}}(1-\eta_1\eta_2^*)^{-\ell-1},
\end{equation}
and satisfy the completeness relation
\begin{eqnarray}
	\int |\psi_{\eta}^\ell\rangle\langle \psi_{\eta}^\ell|d\mu(\eta,\ell)=\sum_{n=0}^\infty|\psi_n^{\ell}\rangle\langle \psi_n^{\ell}|=\mathbb{I}
\end{eqnarray}
where the measure $d\mu(\eta,\ell)=\frac{\ell}{\pi}\frac{d^2\eta}{(1-|\eta|^2)^2}$.\\
As we noted for the Barut-Girardello coherent states,  for any $|\Psi\rangle=\sum_{n=0}^\infty c_n|\psi_n^{\ell}\rangle$
in the Hilbert space, one can construct  an analytic function
\begin{eqnarray}
	f(\eta)=(1-|\eta|^2)^{-\ell-1}\langle \psi_{\eta}^\ell|\Psi\rangle=\sum_{n=0}^\infty c_n
	\sqrt{\frac{\Gamma(n+\ell+1)}{n!\Gamma(\ell+1)}}(\eta^*)^n.
\end{eqnarray}
The expansion of $|\Psi\rangle$ on the bases of coherent states (\ref{lan36}) can be written as 
\begin{eqnarray}
	|\Psi\rangle &=&\int d\mu(\eta,\ell)(1-|\eta|^2)^{\frac{\ell+1}{2}}f(\eta)|\psi_{\eta}^\ell\rangle,\\
	\langle\Psi|\Psi\rangle&=&\int d\mu(\eta,\ell)(1-|\eta|^2)^{\ell+1}|f(\eta)|^2<\infty.
\end{eqnarray}
\space\\

We have reported in this chapter  a Landau particle in time-dependent background electric field with time-dependent mass and frequency. 
We have studied this system at the classical level and  have formulated  the corresponding quantum system. At the classical
level we solved the equations of motion and  at  the quantum level, we used the Lewis-Riesenfeld’s method to construct the spectra
of the invariant operator $I(t)$ and the Hamiltonian $H(t)$ on the helicity-like bases $|\phi_{n\pm}(t)\rangle$.
The configuration space wave functions of both operators are expressed in terms of the generalised Laguerre polynomials.
This quantization has shown  that the system possesses $U(1), SU(2)$ and $SU(1,1)$ symmetries. Consequently,  a system  of 
coherent states associated to those symmetries  is constructed and  some usual  related  properties  to those states are examined.

 \section*{Acknowledgements}
 This  topic is part of my PhD thesis  entitled {\it ''Generalized Uncertainty Principle and Time-dependent Landau Problem''} that has been  defended at IMSP. I would like to take  this opportunity to renew my thank to my supervisor Prof Gabriel Avossevou who  have provided me this topic, advice and  moral support during this periode of study. I would like to express also my deepest gratitude to Dr Laure Gouba for having accepted to collaborate with us on  my thesis. Thanks to  the lecturers of IMSP for their excellent guidance, advice and encouragements during my doctoral research courses. I am deeply indebted to The Abdus Salam International Centre for Theoretical Physics (ICTP) and the German Academic Exchange Service (DAAD) for their  financial supports. Finally, I am greatly indebted to my family for their love and support to me. My parents and
 brother have always been my best supporters. I deeply thank God who has led me so far.


\begin{thebibliography}{99}
 	\bibitem{102}    L. Lawson and G. Avossevou, J. Math. Phys.  {\bfseries 59}, 042109 (2018)
 
 
  	\bibitem{178}    J. Govaerts,  M. Hounkonnou and H. Mweene,  {\bfseries 42}, 485209 (2009)
  	\bibitem{179}    J. Geloun, J. Govaerts and F. Scholtz, J. Phys. A: Math. Theor. {\bfseries 42}, 495203 (2009)
  	\bibitem{180}    J. Govaerts, M. Hounkonnou and H. Mweene,  J. Phys A: Math.Theor {\bfseries 42}, 485209 (2009)
  	\bibitem{132}   V. Ermakov   Univ. Izv. Kiev {\bfseries 20}, 1  (1880)
  	
  	\bibitem{133}   V. Ermakov   Appl. Anal. Discrete Math. {\bfseries 2}, 123 (2008)
  	 	
  	\bibitem{134}   W. Milne  Phys. Rev. {\bfseries 35}, 86367 (1930)
 
 		\bibitem{135}   E. Pinney  Proc. A.M.S. {\bfseries 1}, 681  (1950)
 		
 		 	\bibitem{104}    G. Fiore and L. Gouba, J. Math. Phys. {\bfseries 52}, 103509 (2011)
 		 	\bibitem{105}    M. Maamache, A. Bounames  and N. Ferkous, Phys Rev A {\bfseries 73}, 016101 (2006)
 		 	\bibitem{106}    Y. Bouguerra, M. Maamache and A. Bounames,  Int. J. Theor Phys {\bfseries 45}, 9 (2006)
 		 	\bibitem{107}    C. Ferreira, P. Alencar and J. Bassalo, Phys Rev A {\bfseries 66}, 024103 (2002)
 		
 		 	\bibitem{181}    M. Maamache, A. Bounames  and N. Ferkous, Phys Rev A {\bfseries 73}, 016101 (2006)
 		 	\bibitem{182}    Y. Bouguerra, M. Maamache and A. Bounames,  Int. J. Theor Phys {\bfseries 45}, 9 (2006)
 		 	\bibitem{183}    C. Ferreira, P. Alencar and J. Bassalo, Phys Rev A {\bfseries 66}, 024103 (2002)
 		 	
 		 \bibitem{47}  H. Lewis  and  W. Riesenfeld, J. Math. Phys. {\bfseries 10}, 1458-1473  (1969)
 		 	\bibitem{47'}    L.Lawson,G. Avossevou and L. Gouba J. Math. Phys. {\bfseries 59}, 112101 (2018)
 		 
 		 	\bibitem{84}     O. Barut and L. Girardello, Commun. Math. Phys {\bfseries 21}, 41 (1971)
 		 	
 		 		\bibitem{169}      I. Gradshteyn and M. Ryzhik: Table of Integrals, Series, and Products, Academic Press, 
 		 	 	San Diego, Calif, USA, 8th edition, (2015)
 		 	 		\bibitem{83}     A. Perelomov, Commun. Math. Phys {\bfseries 26}, 222-236 (1972)
 		
 		
 		
%
%
%
%
%
%
%
%
%

%
%
%
%
%

%
%
%
%
%
%

%
%
%
%
%
%
%
%
%

%
%
%
 	
 	
 \end{thebibliography}
\end{document}